\documentclass[a4paper,11pt]{article}
\pdfoutput=1 % if your are submitting a pdflatex (i.e. if you have
\usepackage{jheppub} % for details on the use of the package, please
\usepackage[T1]{fontenc} % if needed

\usepackage[english]{babel}

\usepackage{bm,amssymb,amsmath,slashed,graphicx,multirow,soul,mathtools,xspace,array,epsfig,color,bm}
\usepackage{siunitx}

\usepackage{subcaption}

\graphicspath{{Figures/}} 

\title{Minimally modified Fritzsch texture for quark masses and CKM mixing}

\author[a,b]{Benedetta Belfatto,}
\author[c,d]{Zurab Berezhiani}

\affiliation[a]{SISSA International School for Advanced Studies, \\
  Via Bonomea 265, 34136, Trieste, Italy}
\affiliation[b]{INFN - Sezione di Trieste, \\
Via Bonomea 265, 34136, Trieste, Italy}
\affiliation[c]{Dipartimento di Scienze Fisiche e Chimiche, Universit\`a di L'Aquila,\\ 67100 Coppito, L'Aquila, Italy}
\affiliation[d]{INFN, Laboratori Nazionali del Gran Sasso,\\ 67010 Assergi, L'Aquila, Italy}

\emailAdd{bbelfatt@sissa.it}
\emailAdd{zurab.berezhiani@aquila.infn.it}

\abstract{The Standard Model does not constrain the form of the Yukawa matrices and thus
the origin of fermion mass hierarchies and mixing pattern remains puzzling.
On the other hand, there are intriguing relations between fermion masses and mixing angles
which may point towards specific textures of Yukawa matrices.
One of the classic hypothesis is the zero texture proposed by Fritzsch
which is, however, excluded by present precision tests since it
predicts a too large value of $|V_{cb}|$ as well as a too small value of the ratio $|V_{ub}/V_{cb}|$.
In this paper we discuss a minimal modification which still maintains the six
zero entries as in the original Fritzsch ansatz.
This modification consists in introducing an asymmetry between the 23 and 32 entries
in the down-quark Yukawa matrix.  We show that this flavour structure can naturally emerge in the context of
models with inter-family $SU(3)_H$ symmetry.
We present a detailed analysis of this Fritzsch-like texture by testing its
predictions and showing that it is perfectly compatible with the present precision data
on quark masses and CKM mixing matrix. 
}

\begin{document} 
\maketitle
\flushbottom

\section{Introduction}

The replication of fermion families is one of the main puzzles of particle physics. 
Three fermion families are in identical representations of the Standard Model (SM) 
gauge symmetry $SU(3)\times SU(2)\times U(1)$.
Left-handed quarks $q_{Li}=(u_L,d_L)_i$ and leptons $\ell_{Li}=(\nu_L,e_L)_i$  
transform as weak doublets whereas right-handed components 
$u_{Ri},d_{Ri},e_{Ri}$ are weak singlets, $i=1,2,3$ being the family index. 
The fermion masses emerge after spontaneous breaking of the EW symmetry 
$SU(2)\times U(1)$ by the Higgs doublet $\phi$, via the Yukawa couplings 
\begin{align}
\label{Yukawas-SM}
 Y_u^{ij}   u^c_i q_j  \phi  + Y_d^{ij}  d^c_i q_j \tilde\phi   + Y_e^{ij}  e^c_i \ell_j \tilde\phi
 \, + \, {\rm h.c.}  
\end{align}
where $Y_{e,u,d}$ are the Yukawa matrices, and $\tilde\phi=i\tau_2 \phi^\ast$.  
Here, instead of the right-handed fermion fields, we use their left-handed complex conjugates
(antifields) as $u^c_{L}= {\rm C} \overline{u_{R}}^T$ and omit in the following 
the subscript $L$ for  $q$, $u^c$, $d^c$ etc. all being the left-handed Weyl spinors. 
With these notations, the description can be conveniently extended to a 
supersymmetric extensions of the SM and/or to a grand unified theory (GUT). 
The Yukawa couplings \eqref{Yukawas-SM},  after substituting the Higgs vacuum 
expectation value (VEV) $\langle \phi^0 \rangle = v_{\rm w} = 174$ GeV, 
originate the fermion mass matrices $\mathrm{M}_f = Y_f v_{\rm w}$, $f = u,d,e$.  
 They can be brought to the diagonal form (the mass eigenstate basis) 
 via the bi-unitary transformations: 
\begin{align}
& V_{f}^c  \mathrm{M}_f V_{f} = \mathrm{M}_f^{\rm diag} 
\label{diag}
\end{align}
so that the quark masses  $m_u,m_c,m_t$ and $m_d,m_s,m_b$   
are the eigenvalues of the mass matrices $\mathrm{M}_u$ and $\mathrm{M}_d$. 
(In the following we discuss concretely the quark sector considering  
the presence of leptons implicitly.) 
The ``right" matrices $V_{u,d}^c$ rotating the right-handed quarks 
have no physical meaning in the SM context, 
while the ``left" ones $V_{u,d}$ give rise to the 
mixing
in the quark charged currents coupled to weak $W^\pm$ bosons
which is determined by the unitary Cabibbo-Kobayashi-Maskawa (CKM) mixing matrix $V_\text{CKM}$
\cite{Cabibbo:1963yz,Kobayashi:1973fv}:
\begin{align}\label{CKM}
V_\text{CKM}= V_{u}^\dagger V^{\,}_{d}= \left(\begin{array}{ccc}
V_{ud} & V_{us} & V_{ub}  \\
V_{cd} & V_{cs} & V_{cb}  \\
V_{td} & V_{ts} & V_{tb} 
\end{array}\right) 
\end{align}
 This matrix is unitary, and by rotating away the irrelevant phases, 
it  can be conveniently parameterized in terms of four parameters, 
 three mixing angles $\theta_{12}$, $\theta_{23}$,  $\theta_{13}$ and one CP-violating phase $\delta$
 \cite{Kobayashi:1973fv}.
 In the so called standard parameterization \cite{PDG22}  
 it is written as:
 \begin{align}
 \label{ckmSP}
V_\text{CKM}=  \left(\begin{array}{ccc}
c_{12}c_{13} & s_{12}c_{13} & s_{13}e^{-i\delta} \\
-s_{12}c_{23}-c_{12}s_{23}s_{13}e^{i\delta} & c_{12}c_{23}-s_{12}s_{23}s_{13}e^{i\delta} 
& s_{23}c_{13}  \\
s_{12}s_{23}-c_{12}c_{23}s_{13}e^{i\delta} & -c_{12}s_{23}-s_{12}c_{23}s_{13}e^{i\delta} 
& c_{23}c_{13}
\end{array}\right) 
\end{align}
where $s_{ij}=\sin\theta_{ij}$, $c_{ij}=\cos\theta_{ij}$ and $\theta_{ij}$ 
can be chosen so that $s_{ij},c_{ij}\geq 0$.
As a measure of CP violation, the rephasing-invariant quantity
$J\sum_{m,n}\epsilon_{ikm}\epsilon_{jln}=\text{Im}[V_{ij}V_{kl}V_{il}^*V_{kj}^*]$ can be considered
instead of the phase $\delta$,
the Jarlskog invariant \cite{Jarlskog:1985ht}, which in the standard parameterization reads:
\begin{align}
& J=\sin\delta\, s_{12}s_{23}s_{13} \, c_{12}c_{23}c_{13}^2 
\end{align}

The mass spectrum and the mixing angles of quarks present a strong inter-family hierarchy.
Namely, by parameterizing masses and mixings between quarks
with a small parameter $\epsilon \sim1/20$, we have 
for down-type quarks $m_b:m_s:m_d=1:\epsilon:\epsilon^2$ and for up-type
$m_t:m_c:m_u=1:\epsilon^2:\epsilon^4$,
with $V_{us}\sim \sqrt{\epsilon}$, $V_{cb}\sim \epsilon$, $V_{ub}\sim \epsilon^2$.
The SM does not contain any theoretical input that could explain 
the inter-family hierarchy of fermion masses and the pattern of the CKM mixing angles. 
Besides, the same is true for its supersymmetric or grand unified extensions. 
In a sense, the SM is technically natural since it can tolerate any  pattern
of the Yukawa matrices   $Y_f$, 
but it tells nothing about  their  structures which remain arbitrary.       
So the origin of the fermion mass hierarchy and their weak mixing pattern 
remains a mystery.  

It is tempting to think that 
the fermion flavour structure is connected to some underlying theory which determines
 the pattern of the Yukawa matrices, and that
relations between masses and mixing angles such as the
well-known formula for the Cabibbo angle $V_{us} = \sqrt{m_d/m_s}$ are not accidental.
In particular, relations between the fermion masses and mixing angles can be obtained by considering Yukawa matrix textures with reduced number of free parameters, with certain zero elements. 
This {\it zero-texture} approach was originally thought to calculate the Cabibbo angle in the 
two-family framework in refs.~\cite{Weinberg1977,Wilczek1977,Fritzsch1977}, 
in fact before the discovery of $b$ and $t$ quarks. 

In the frame of six quarks, in refs. \cite{Fritzsch78,Fritzsch79} H. Fritzsch 
extended the zero texture for the mass matrices 
in the form:
\begin{align}\label{fritzsch}
\mathrm{Y}_{u,d}=\left(\begin{array}{ccc}
0 & A'_{u,d} & 0 \\ A_{u,d} &0 & B'_{u,d} \\ 0 & B_{u,d} & C_{u,d}
\end{array}\right) 
\end{align}
where all non-zero elements are generically complex, with the symmetricity condition 
$|A_{u,d}|=|A'_{u,d}|$,  $|B_{u,d}|=|B'_{u,d}|$
motivated in the context of the left-right symmetric models. 
Besides reproducing the formula for the Cabibbo angle, 
this texture exhibits at least two remarkable features.
\begin{itemize}
\item
By a phase transformation of the quark fields, 
$F'_{u,d} \mathrm{M}_{u,d} F_{u,d}=\widetilde{\mathrm{M}}_{u,d}$,  the matrices \eqref{fritzsch} 
can be brought to real symmetric matrices $\widetilde{\mathrm{M}}_{u,d}$, which then can 
be diagonalized by orthogonal transformations,  $O_{u,d}^T \widetilde{\mathrm{M}}_{u,d} O_{u,d}
= \mathrm{M}^{\rm diag}_{u,d}$.  
In this way, the three real parameters $|A_d|$,  $|B_d|$, $|C_d|$ can be expressed 
in terms of the three eigenvalues of $\widetilde{\mathrm{M}}_d$, i.e. the down quark masses 
$m_d,m_s,m_b$, and so the three rotation angles in the orthogonal matrix $O_d$ 
can be expressed in terms of the mass ratios $m_d/m_s$ and $m_s/m_b$. 
Analogously, the three angles in $O_u$ 
can be expressed in terms of the upper quarks mass ratios $m_u/m_c$ and $m_c/m_t$. 
The CKM matrix \eqref{CKM} is obtained as $V_\text{CKM} = O_u^T F_u^\ast F_d O_d$, 
where the diagonal matrix $F=F_u^\ast F_d$ can be parameterized by two phase parameters,  
$F=\text{diag}(e^{i\alpha}, e^{i\beta}, 1)$. Then, the four physical elements 
of the CKM matrix, that is the three mixing angles $\theta_{12},\theta_{23},\theta_{13}$ and the CP-phase $\delta$,  
can be expressed in terms of the known mass ratios, 
$m_d/m_s$, $m_s/m_b$, $m_u/m_c$ and $m_c/m_t$, and of two unknown phases 
$\alpha$ and $\beta$. 
\item In view of the interfamily hierarchies, $m_d \ll m_s \ll m_b$ and  $m_u \ll m_c \ll m_t$, 
the Fritzsch ansatz \eqref{fritzsch} demonstrates 
an interesting property coined as the {\it decoupling hypothesis} \cite{Fritzsch83}.  
Since the CKM angles depend on the quark mass ratios, 
in the limit in which the masses of the first family vanish, $m_u,m_d \to 0$, the mixing 
of the latter with the heavier families should disappear, i.e. $\theta_{12}, \theta_{13} \to 0$. 
At the next step, in the limit of massless second family, $m_s,m_c \to 0$, 
also the 2-3 mixing should disappear, i.e. $\theta_{23} \to 0$.
\end{itemize}
However, in the original works \cite{Fritzsch78,Fritzsch79}, 
the `zeros' in these matrices were achieved  at the price of introducing several Higgs bi-doublets 
differently transforming under some discrete flavor symmetry.  
This underlying theoretical construction looks rather obsolete. Namely, the need for several 
Higgs bi-doublets spoils the natural flavor conservation \cite{GIM,Weinberg,Paschos} 
and unavoidably leads to severe flavor-changing effects \cite{Gatto}.  
In a more natural way, 
without employing the left-right symmetry, the Fritzsch texture can be obtained in the context of models 
with $SU(3)_H$ gauge symmetry between the three families \cite{PLB,PLB2}, as we shall describe 
in this work. 

Moreover, 
in light of present experimental and lattice results on quark masses and CKM elements,
the symmetric Fritzsch texture for quarks must be excluded, since there is no parameter 
space in which these precise data  
can be reproduced (see refs. \cite{Kang,Ramond:1993kv}).
More concretely, the small enough value of $|V_{cb}|$ and large enough value of $|V_{ub}/V_{cb}|$ 
cannot be achieved for any values of the phase parameters. 
A possibility to obtain viable textures  
is to extend the original Fritzsch texture by replacing one of the zero entries 
with a non-zero one, e.g. by introducing a non-zero 13 element \cite{Lavoura} 
or a non-zero 22 element, as recently analysed e.g. in refs. \cite{Giraldo:2011ya,Xing:2015sva,Linster:2018avp,Bagai,Fritzsch2021},
without renouncing to the `symmetricity'. 
However, these modifications do not satisfy the {\it decoupling} feature and the introduction 
of new parameters reduces the predictivity.

On the other hand, instead of decreasing the number of zero entries,
one can 
think to break the symmetricity condition.
Namely, an asymmetry in the 23 blocks of the Yukawa matrices of the form in eq. \eqref{fritzsch} can be introduced,
$|B_d| \neq |B'_d|$ \cite{Rossi1,Roberts:2001zy}. It is worth noting that 
in this scenario the properties of the original texture are preserved.
In fact, the decoupling feature does not require the equality of the moduli of non-diagonal elements in \eqref{fritzsch}.
We will give a detailed study of such minimally modified Fritzsch ansatz.

The paper is organized as follows.
In section \ref{model} we describe how the Fritzsch texture can be obtained within the context of 
the inter-family gauge group $SU(3)_H $, and how it can be minimally deformed in the 2-3 blocks 
in presence of a scalar field in adjoint (octet) representation of $SU(3)_H$. 
In section \ref{texture} we analyse the relations between the parameters of the asymmetric Fritzsch texture
and the quark mass ratios and mixing angles.
In section \ref{analysis} 
we confront the minimally modified Fritzsch matrices with the recent high precision 
determinations of quark masses and CKM matrix elements, 
and show that this flavour structure predicts 
all the masses, the mixing angles and the CP-violating phase in perfect agreement 
with the experimental results.
In section \ref{conclusion} we summarize our results. 

\section{Fritzsch-like textures from horizontal symmetry $\mathbf{SU(3)_H} $ }
\label{model}

The key for understanding the replication of families, fermion mass hierarchy and mixing pattern 
may lie in symmetry principles. 
For example, one can assign to fermion species different charges of an abelian global flavor symmetry  
$U(1)$  \cite{Froggatt}.  
There are also models making use of an anomalous gauge symmetry $U(1)_A$ 
to explain the fermion mass hierarchy while also tackling other naturalness issues 
\cite{Ibanez,Binetruy,Dudas,Tavartkiladze1,Tavartkiladze2}. 
Abelian flavour symmetries with extra Higgs doublets 
have been used to generate Yukawa matrices with vanishing entries
\cite{Grimus:2004hf,Ferreira:2010ir,Serodio:2013gka,Bjorkeroth:2018ipq}.
However, it is difficult to obtain the highly predictive
quark mass matrices with six texture zeros within this approach. 

Nonetheless, one can point to a more complete picture by introducing the non-abelian horizontal gauge symmetry 
$SU(3)_H$ between three families  \cite{PLB,PLB2,Chkareuli,Chkareuli2,Chkareuli3,Berezhiani:1996kk,Berezhiani:2000cg,Belfatto:2018cfo}.   
This symmetry should have a chiral character, 
with the left-handed and right-handed components of quarks (and leptons) transforming 
in different representations of the family symmetry, namely as $SU(3)_H$ 
triplets and anti-triplets respectively, 
so that the fermions cannot acquire masses without the breakdown of $SU(3)_H$ invariance. 
In our chiral notations this means that all left-handed fields must transform 
as triplets:
\begin{align}\label{SM-triplets}
q_i, \, u^c_i , d^c_i \sim 3 \quad \quad (\ell_i, \, e^c_i \sim 3)
\end{align} 
where $i=1,2,3$ is the family $SU(3)_H$ index. Such an arrangement is compatible 
with the grand unified extensions of the SM. In particular, in the context of $SU(5)$ GUT \cite{Georgi} 
 each family is represented by the left-handed Weyl fields 
 $\bar F_i = (d^c,\ell)_i$ and $T_i = (u^c,e^c,q)_i$ respectively in $\bar5$ and $10$ representations of $SU(5)$. 
Then, the fermions can be arranged in the following representations of 
$SU(5)\times SU(3)_H$ \cite{PLB2,Chkareuli,Chkareuli2}:  
\begin{align}\label{SU5-triplets}
\bar{F}_i = (d^c, \, \ell)_i \sim (\bar5,3) , \quad \quad T_i = (u^c, \, q, \, e^c)_i \sim (10,3) 
\end{align} 
while in the context of $SO(10)\times SU(3)_H$ all these fermions, along with 
the ``right-handed neutrinos" $\nu^c_L=C\overline{\nu_R}^T$ can be packed 
into the unique multiplet in the spinor representation of $SO(10)$, 
$\Psi_i = (\bar{F}, T, \nu^c)_i \sim (16,3)$.\footnote{With this set of fermions,
$SU(3)_H$ would have triangle anomalies. 
For their cancellation one can introduce additional 
 chiral fermions transforming under $SU(3)_H$ \cite{PLB,PLB2,Chkareuli,Chkareuli2}. 
The easiest way to cancel the anomalies is to share the $SU(3)_H$ symmetry 
with mirror fermions \cite{Berezhiani:1996ii}
belonging to a parallel SM$'$ sector of particles identical to the $SM$ 
sector of ordinary particles (for a review, see e.g. \cite{IJMPA,Alice}).  }

Due to the chiral character of the horizontal symmetry, 
the fermion masses cannot be induced without  breaking $SU(3)_H$,
which forbids the direct Yukawa couplings of fermions \eqref{SM-triplets} with the Higgs doublets $\phi$.  
As far as the fermion bilinears $u^c_i q_j$, $d^c_i q_j$ and $e^c \ell$ 
transform in representations $3\times 3= 6 + \bar3$,  
the fermion masses can be induced only via the higher order operators 
\begin{align}\label{operators}
\frac{\, \chi^{ij} }{M} \, u^c_i q_j  \phi \, + \, \frac{\, \chi^{ij} }{M} \, d^c_i q_j \tilde\phi  
\, + \,  \frac{\, \chi^{ij} }{M} \, e^c_i \ell_j \tilde\phi  ~ + ~ \text{h.c.}
\end{align}
involving some horizontal scalars $\chi$ (coined as flavons) in symmetric 
(anti-sextets $\chi^{\{ij\}} \sim \bar6$) or antisymmetric 
(triplets $\chi^{[ij]} =\epsilon^{ijk}\chi_k \sim 3$) representations of $SU(3)_H$, 
where $M$ is some effective scale 
(the coupling constants of different flavons are omitted). 
After inserting the flavon VEVs  in the operators \eqref{operators}, 
the standard Yukawa couplings  \eqref{Yukawas-SM} are induced 
which will reflect the VEVs pattern.  
Extending the SM to $SU(5)$ GUT, 
in the context of $SU(5)\times SU(3)_H$ theory \cite{PLB2,Chkareuli}, 
the Yukawa couplings 
 emerge from the decomposition of the $SU(5)$-invariant Yukawa couplings
\begin{align}
\label{Yukawas-SU5}
 G_u^{ij}  T_i T_j  H   + G_d^{ij}  \bar{F}_i T_j H^\ast  \, + \, {\rm h.c.}  
\end{align}
where $H$ is the scalar $5$-plet which contains the SM Higgs doublet $\phi$. 
$ G_u^{ij}$ and $G_d^{ij}$ are effective Wilson coefficients of operators containing flavons,
emerging from the structures $(\chi^{ij}/M) T_i T_j H$ and  $(\chi^{ij}/M) \bar{F}_i T_j H^\ast$. % where  
Some $\chi$-flavons can also be in adjoint representations of $SU(5)$,
or more generally these effective coefficients should involve a scalar 24-plet
$\Sigma$ of $SU(5)$ %$G_{u,d}=G_{u,d}(\Sigma/M)$,
in order to avoid the undesiderable relations between the down quark and lepton masses \cite{PLB2}.\footnote{
The minimal scenario, with matrices $G_u$ and $G_d$ being $SU(5)$ singlets, would imply   
$Y_u=Y_u^T$ and $Y_e=Y_d^T$, the latter equality leading to incorrect relations 
between the down quark and charged lepton masses. However, this shortcoming can be avoided 
in a more general context, 
by considering $G_{u,d}=G_{u,d}(\Sigma/M)$ as functions of the scalar $\Sigma$ 
in the adjoint representation (24-plet) which breaks $SU(5)$ down to the SM gauge group 
$SU(3)\times SU(2)\times U(1)$ 
at the GUT scale $M_G\simeq 10^{16}$ GeV or so. This is equivalent to introducing 
higher order operators in powers of $\Sigma/M$,  which can be obtained e.g. 
by integrating out some heavy vector-like fermions at the mass scale $M > M_G$. 
In this way, the expansions 
$G_{u,d}(\Sigma)=G^{(0)}_{u,d} + G^{(1)}_{u,d}(\Sigma/M) + G^{(2)}_{u,d}(\Sigma^2/M^2) + \dots $ 
will in general contain terms in 1, 24 etc. representations of $SU(5)$ which remove the 
above restrictive relations $Y_u=Y_u^T$ and $Y_e=Y_d^T$ and render 
the Yukawa matrices $Y_{u,d,e}$ in the low energy SM to be independent from each other
(for a review, see e.g. \cite{ICTP}). }
In the following, 
we mainly concentrate 
on the quark sector in the context of the SM, having in mind that in the context of grand unification
analogous considerations can be extended to leptons.

Interestingly, operators \eqref{operators} which are invariant under the local 
$SU(3)_H$ symmetry by construction, in fact have a larger global symmetry $U(3)_H$. 
Namely,  they are invariant also under a global chiral $U(1)_H$ symmetry,  
implying an overall phase transformation of fermions $u^c_i, d^c_i,q_i$ 
and flavon scalars $\chi$.
Hence, all families can become massive only if $U(3)_H$ symmetry is fully broken. 

 This feature allows to relate the fermion mass hierarchy and mixing pattern with the 
breaking pattern of $U(3)_H$ symmetry, with a natural realization of the decoupling hypothesis.  
When $U(3)_H$ breaks down to $U(2)_H$, the third family of fermions become massive
while the first two families remain massless, and mixing angles are zero.
At the next step, when $U(2)_H$ breaks down to $U(1)_H$,  
the second family acquires masses and the CKM mixing angle $\theta_{23}$ can be non-zero, 
but the first family remains massless ($m_u,m_d=0$) and unmixed with the heavier 
fermions  ($\theta_{12},\theta_{13}=0$). Only at the last step, when $U(1)_H$ is broken, 
also the first family can acquire masses and its mixing with heavier families can emerge. 
In this way, the inter-family mass hierarchy can be related to the hierarchy  
of flavon VEVs inducing the horizontal symmetry breaking 
$U(3)_H\to U(2)_H \to U(1)_H \to$ {\it nothing}.

In the last step of this breaking chain, the chiral global $U(1)_H$ symmetry can be associated with the 
Peccei-Quinn symmetry provided that $U(1)_H$ is also respected by the Lagrangian 
of the flavon fields \cite{PLB2,Khlopov1}. This can be achieved by forbidding 
the trilinear terms between the $\chi$-scalars by means of a discrete symmetry. 
Thus, in this framework, the Peccei-Quinn symmetry can be considered as an accidental symmetry 
emerging from the local symmetry $SU(3)_H$. 
In this case the axion will have non-diagonal couplings between  
the fermions of different families, i.e. it will act as a familon \cite{PLB,Khlopov1}.  
Phenomenological and cosmological implications of such flavor-changing axion 
were discussed in refs. 
\cite{Khlopov2,Khlopov3,Khlopov4,Homeriki1,Homeriki2,Sakharov,DiLuzio:2017ogq,DiLuzio:2019mie,MartinCamalich:2020dfe,Calibbi:2020jvd}.

Let us discuss now how Fritzsch zero textures can naturally emerge in this scenario with horizontal symmetry.  
As the simplest set of $\chi$-flavons, we can choose two triplets $\chi_1$, $\chi_2$,
and one anti-sextet $\chi_3$,  and arrange their VEVs in the following form \cite{PLB2}: 
\begin{align}\label{vevs}
&\langle\chi^{\{ij\}}_3 \rangle=\text{diag}(0,0,V_3) \qquad
\langle\chi_{2i} \rangle=\left(\begin{array}{c} V_2 \\ 0 \\0 \end{array}\right) \qquad
\langle\chi_{1i}\rangle=\left(\begin{array}{c} 0 \\ 0 \\V_1 \end{array}\right)
\end{align}
i.e. the VEV 
of $\chi_3 $ is given by a symmetric rang-1  matrix 
directed towards the 3rd axis in the $SU(3)_H$ space,
the VEV of $\chi_1$ is parallel to $\langle \chi_3 \rangle$ and 
the VEV of $\xi$ is orthogonal to it and without losing generality 
it can be oriented towards the 1st axis (for the analysis of the flavon potential 
allowing such a solution see ref.~\cite{Chkareuli}). 
The total matrix of flavon VEVs has the form 
\begin{align}\label{VEV-matrix}
\langle \chi^{ij} \rangle = \left\langle \chi_1^{[ij]} + \chi_2^{[ij]} + \chi_3^{\{ij\}} \right\rangle= 
  \left(\begin{array}{ccc} 0 &  V_1 & 0 \\ -  V_1 & 0 &   V_2 \\ 0 & -  V_2 &  V_3 
  \end{array}\right)
\end{align}
Then, modulo different coupling constants 
of $\chi$-flavons in the two operators in \eqref{operators}, the Yukawa matrices 
$Y_u, Y_d \propto \langle \chi \rangle/M$ will
reflect the VEV pattern \eqref{VEV-matrix}. Hence,  the Yukawa matrices would acquire
the `symmetric' Fritzsch forms \eqref{fritzsch} with $A'_{u,d} = A_{u,d}$ 
and  $B'_{u,d} = B_{u,d}$ (the $-$ signs can be eliminated by quark phase transformations).  
The hierarchies between the different Yukawa entries, corresponding to the
inter-family mass hierarchies, can be related to a hierarchy 
$V_3 \gg V_2 \gg V_1$ in the horizontal symmetry breaking chain 
$U(3)_H\to U(2)_H \to U(1)_H \to$ {\it nothing}. 
After this breaking, the theory reduces to the SM with one standard Higgs doublet $\phi$, 
and so, in difference from the Fritzsch's original model \cite{Fritzsch78,Fritzsch79}, 
in our construction the flavor will be naturally conserved in neutral currents 
\cite{GIM,Weinberg,Paschos}. 

In the UV-complete pictures the operators \eqref{operators}  
can be induced via renormalizable interactions after integrating out  some extra heavy fields, 
scalars \cite{Chkareuli} or verctor-like fermions \cite{PLB,PLB2,Khlopov1}. 
Hereafter we shall employ the second possibility. 
Namely,  one can introduce the following set of left-handed fermions of up- and down-quark type 
in weak singlet representations
 \begin{align}\label{VL}
U_i, \, D_i \sim 3 \qquad U^{ci},\, D^{ci} \sim \bar3 
\end{align} 
These fermions are allowed to have $SU(3)_H$ invariant mass terms.  
More generically, their  mass terms transform as $\bar3\times 3=1+8$ and 
they can emerge from the Yukawa couplings 
with the scalars in singlet and octet representations of 
$SU(3)_H$, $S \sim 1$ and $\Phi \sim 8$, namely 
\begin{align}\label{Yuk-Phi} 
(g_D S \delta_j^i +  f_D \Phi_j^i) D_i D^{cj} + {\rm h.c.} 
\end{align}
with analogous couplings for $U_i, U^{ci}$. 
In fact,  one can introduce an adjoint scalar $\Phi$ of $SU(3)_H$ 
in analogy to the adjoint scalar of $SU(5)$, the 24-plet $\Sigma$.
We also assume 
that the cross-interaction terms of $\Phi$ with $\chi$-flavons in the scalar potential 
align the $\Phi$ VEV towards the largest VEV $V_3$ in \eqref{VEV-matrix}, i.e. proportionally to the $\lambda_3$ 
generator: $\langle \Phi \rangle = V \, \mathrm{diag}(1,1,-2)$.  
In this case, the heavy fermion mass matrices contributed by singlet and octet VEVs 
have the diagonal  forms 
\begin{align}\label{MUD}
\mathrm{M}_{U,D} = g_{U,D} \langle S\rangle + f_{U,D} \langle \Phi \rangle = 
M_{U,D} \, \mathrm{diag} (X_{U,D}^{-1},X_{U,D}^{-1},1)
\end{align}  
where $M_{U,D}\sim M$ is an overall mass scale determined by the VEVs  of $S$ and $\Phi$ 
and generically $X_{U,D} \neq 1$ are complex numbers. 
Only in the absence of the octet contribution we have $X_{U,D}=1$.

\begin{figure}
\centering
\includegraphics[width=0.48\linewidth]{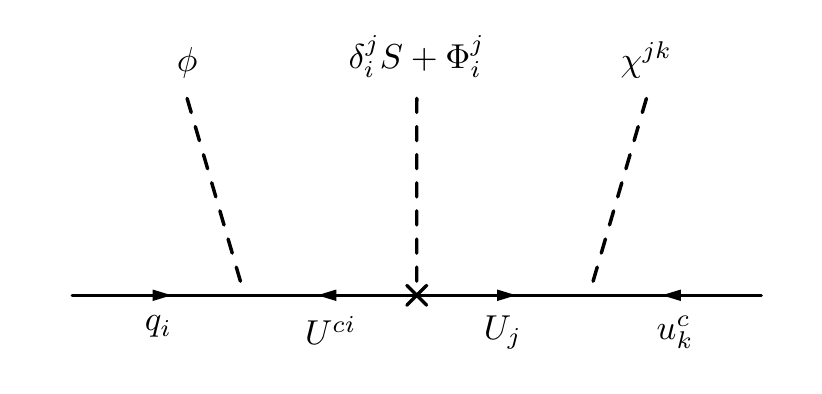}
\hspace{10pt}
\includegraphics[width=0.48\linewidth]{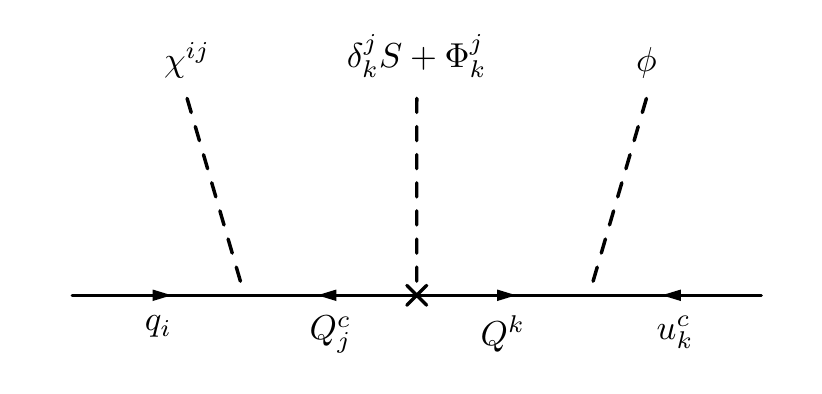}
\caption{\label{massdiag} Seesaw diagrams inducing the Yukawa couplings of upper quarks 
via exchange of vector-like quarks $U,U^c$ and $Q,Q^c$. 
Analogous diagrams with $U,U^c\to D,D^c$ and $\phi \to \tilde\phi$ will work for down quarks.
}
\end{figure}

The following Yukawa couplings between the light quarks $q_i, u^c_i, d^c_i \sim 3$ \eqref{SM-triplets}
and heavy fermions \eqref{VL} are allowed by the symmetry 
\begin{align}\label{Yukawas}
\sum_{n=1}^3 h_{u}^{(n)} \chi_n^{ij} \, u^c_i U_j 
+ \sum_{n=1}^3 h_{d}^{(n)} \chi_n^{ij} \, d^c_i D_j 
+ y_u  \phi   \, U^{ci} q_i + y_d \tilde\phi \, D^{ci} q_i 
\end{align} 
with couplings $h_{u,d}$ and $y_{u,d}$. 
In this way, the matrices of Yukawa couplings $Y_{u,d,}$  \eqref{Yukawas} are induced 
after integrating out the heavy fermions via universal seesaw mechanism \cite{PLB,PLB2}.  
Namely, for the upper quarks this mechanism is illustrated by the first diagram 
in figure \ref{massdiag} while the analogous diagram involving $D$ and $D^c$ states 
will work for the down quarks. 
More generally, 
also heavy quarks in weak doublet representations 
$Q^i\sim \bar3$ and $Q^c_i\sim 3$ can be used for quark mass generation 
(see the second diagram in figure \ref{massdiag}). 
However, this would not affect the final form of the Yukawa matrices \eqref{YF-Real} 
which we shall discuss in this work.\footnote{ 
%The recently observed Cabibbo angle anomalies can be an indication of 
%the presence of vector-like quarks with mass of few TeV mixing with light quarks
Mixing of the light quarks with vector-like quarks with mass of order TeV can be at the origin of 
%has also been studied as an explanation of
the recently observed Cabibbo angle anomalies \cite{Belfatto:2019swo,Belfatto:2021jhf,Cheung:2020vqm,Branco:2021vhs,Botella:2021uxz,Crivellin:2022rhw,Belfatto:2023tbv} (see ref. \cite{Fischer} for a review).}
Analogously, the charged lepton Yukawa couplings 
can be induced by introducing the vector like lepton states, weak singlets 
 $E_i \sim 3$, $E^{ci} \sim \bar3$ and weak doublets $L^i \sim \bar3$, $L^c_i \sim 3$. 
In particular, they will be at work in the case of $SU(5)\times SU(3)_H$ extension \cite{PLB2} 
where all these states fit into the  set of fermions in vector-like representations 
$(\bar5,\bar3) =  (D^{ci}, L^i)$, $(5,3) = (D_i,L^c_i)$ and  
$(10,\bar3) = (U^{ci},Q^i,E^{ci})$, $(\overline{10},3) = (U_i, Q^c_i, E_i)$. 
Interestingly, in the context of supersymmetry 
our mechanism can lead to interesting relations 
between the fermion Yukawa couplings and the soft SUSY breaking terms which allow 
to naturally  realize the minimal flavor violation scenarios 
\cite{Berezhiani:1996ii,MFVA,MFVR,MFVS}.  
After substituting the flavon VEVs in eq. \eqref{Yukawas}, we obtain  
 \begin{align}\label{seesaw}  
 Y_u = \chi_{u} \mathrm{M}_U^{-1}  y_u , \qquad Y_d = \chi_d  \mathrm{M}_D^{-1} y_d; \qquad 
 \qquad  \chi_{u,d}^{ij} = \sum_{n=1}^3 h_{u,d}^{(n)} \langle \chi^{ij}_n \rangle   
 \end{align} 
Therefore, 
the Yukawa couplings of quarks will have the forms
\begin{align}\label{YF-vevs}
&Y_d = \frac{y_d}{M_D} \left(\begin{array}{ccc}
0 & X_D h^{(1)}_D V_1  & 0 \\ 
- X_D h^{(1)}_D V_1 &0 & h^{(2)}_D V_2 \\
 0 & - X_D h^{(2)}_D V_2 & h^{(3)}_D V_3
\end{array}\right)  , \nonumber \\
&Y_u = \frac{y_u}{M_U} \left(\begin{array}{ccc}
0 & X_U h^{(1)}_U V_1  & 0 \\ 
- X_U h^{(1)}_U V_1 &0 & h^{(2)}_U V_2 \\
 0 & - X_U h^{(2)}_D V_2 & h^{(3)}_U V_3
\end{array}\right) 
\end{align}
where the non-zero entries are generically complex. 
By the phase transformations  $\widetilde{Y}_{u,d}  = F^{\prime}_{u,d} Y_{u,d} F_{u,d}$, 
where $F_{d,u} = \mathrm{diag}(e^{i\alpha_{d,u}}, e^{i\beta_{d,u}}, e^{i\gamma_{d,u}})$, 
the Yukawa matrices can be brought to the forms
\begin{align}
&\widetilde{Y}_d =\left(\begin{array}{ccc}
0 & A_d  & 0 \\ A_d  &0 & x_d B_d \\ 0 & 
x_d^{-1}{B_d}  & C_d
\end{array}\right)  , \quad
\widetilde{Y}_u =\left(\begin{array}{ccc}
0 & A_u   & 0 \\ A_u   &0 & x_u B_u  \\ 0 & 
x_u^{-1}{B_u}  & C_u
\end{array}\right) 
\label{YF-Real}
\end{align}
with all parameters being real and positive. 
In the absence of the 
$SU(3)_H$ octet contribution in the heavy fermion masses we would have 
$x_{u,d}=1$ and thus we would effectively obtain the ``symmetric" Fritzsch ansatz \eqref{fritzsch}. 
However, 
this possibility is excluded since it predicts too large value of $\vert V_{cb}\vert $ and too small  value of 
$\vert V_{ub}/V_{cb}\vert$.

Let us conclude this section with the following remark. In the context of $SU(5)\times SU(3)_H$ theory 
there is a symmetry argument which can lead to $x_u=1$ 
for the up-quark Yukawa matrix \eqref{YF-Real}. 
In this case the quark states are packed together with leptons in $(\bar5,3)$ and $(10,3)$ 
multiplets, as in \eqref{SU5-triplets}. 
For the generation of down-quark and lepton masses 
one can introduce the heavy vector-like fermions 
$\overline{\mathcal{F}}^i = (D^c,L)^i$ and $\mathcal{F}_i = (D,L^c)_i$ 
respectively in $(\bar5,\bar3)+(5,3)$ representations, which contain the down-type quarks in eq. \eqref{VL}, 
with Yukawa couplings   
$h_d^{(n)} \chi_n \overline{F} \mathcal{F} + y_d \mathcal{F} T \overline{H}$,
which include the down quark couplings of \eqref{Yukawas} and where $H$ is the Higgs 5-plet of $SU(5)$. 
The heavy fermions can get masses via extending \eqref{Yuk-Phi} with the Yukawa couplings 
\begin{align}\label{SU5-down}
(g_D S +  f_D \Phi + h_D \Sigma) \mathcal{F} \overline{\mathcal{F} } +\mathrm{h.c}
\end{align}
  in which the 24-scalar $\Sigma$ of $SU(5)$ can also participate.  
In this way, the down quark Yukawa matrix have the form \eqref{YF-Real}.
Also the Yukawa matrix of charged leptons will present analogous form,
 with asymmetry parameter $x_e$, while
the degeneracy between the down quark and lepton masses is removed by the contribution 
of the VEV $\langle \Sigma\rangle $ which breaks  $SU(5)$. 

The up-type quarks in eq. \eqref{VL} are contained in the multiplets of vector-like fermions 
$\mathcal{T}^i =(Q,U^c,E^c)^i$ and $\overline{\mathcal{T} }_i =(Q^c,U,E)_i$ 
in representations $(10,\bar3)+(\overline{10},3)$, 
with the couplings
$h_u^{(n)} \chi_n T \overline{\mathcal{T} }  +  y_u \mathcal{T} T H $ inducing the masses of the upper quarks. 
The heavy fermions can get mass via the Yukawa couplings 
\begin{align}\label{SU5-up}
(g_U S +   h_U \Sigma)\overline{\mathcal{T} }  \mathcal{T}  +\mathrm{h.c}
\end{align}
without the contribution of the $SU(3)_H$ octet $\Phi$. Clearly, then the upper quark Yukawas 
are obtained in the form of $Y_u$ in \eqref{YF-vevs} but with $X_U=1$.\footnote{In fact, 
antisymmetric elements 
emerge from the contribution of the effective combination of $\Sigma H$ in representation 
$45 \subset 24 \times 5$. Interestingly, in the context of supersymmetric 
$SU(5)$ theory,  such a structure can also lead 
to natural suppression of dangerous $D=5$ operators \cite{Berezhiani:1998hg}. }
Obviously, by rotating away the fermion phases, this matrix can be reduced 
to $\widetilde{Y}_u$ in \eqref{YF-Real} with $x_u=1$. 

In principle, the vector-like 10-plets can contribute also in down-quark and lepton masses
via the Yukawa term $y'_d \mathcal{T} \overline{F} \overline{H} $, 
and this contributions can also participate in removing the mass degeneracy. % between the latter. 
In fact, by employing symmetry arguments, one can envisage a scenario in which only $S$ and $\Phi$ 
(or only $\Phi$) 
participate in the couplings \eqref{SU5-down}, without being accompanied by $\Sigma$, 
while only $\Sigma$ participates in \eqref{SU5-up}. In this case one can receive
a rather predictive model which, apart of $x_u=1$ and $x_d\neq1$, 
can give specific relations between the quark and lepton masses via the Clebsch structures
of $SU(5)$ and $SU(3)_H$. 
However this consideration goes beyond the purpose of our paper. 
Nevertheless, in the following we shall pay a spacial attention to the case $x_u = 1$ 
in the considered structures for the quark Yukawa matrices \eqref{YF-Real}.

\section{Parameters of the asymmetric Fritzsch texture}
\label{texture}

The Yukawa matrices in the Fritzsch form \eqref{fritzsch} can be diagonalized by biunitary transformations
parameterized as:
\begin{align}
& (F'_{d}O'_{d})^\dag \, Y_d \, (F_{d}O_{d}) =\text{diag}(y_d,y_s,y_b) \, , \qquad
(F'_{u}O'_{u})^\dag \, Y_u \, (F_{u}O_{u})=\text{diag}(y_u,y_c,y_t) 
\end{align}
where $O_{d(u)}$, $O'_{d(u)}$ are orthogonal matrices and
$F_{d(u)}$, $F'_{d(u)}$ are the phase transformations, 
so that the rephased matrices:
\begin{align}
& \widetilde{Y}_d=F_d^{\prime\, \dag} Y_d F_d =\left(\begin{array}{ccc}
0 & A_d & 0 \\
A_d & 0 & B_d\, x_d \\
0 & B_d/x_d & C_d 
\end{array}\right)
\, , \qquad \widetilde{Y}_u=F_u^{\prime\, \dag} Y_u F_u =\left(\begin{array}{ccc}
0 & A_u & 0 \\
A_u & 0 & B_u\, x_u \\
0 & B_u/x_u & C_u 
\end{array}\right)
\label{YFr}
\end{align}
present only real and positive entries.
Then, 
 the real matrices are diagonalized by the bi-orthogonal transformations
\begin{align}\label{Ortho}
& O_d^{\prime T} \widetilde{Y}_d O_{d}=\text{diag}(y_d,y_s,y_b) \, , \qquad
O_u^{\prime T} \widetilde{Y}_u O_{u}=\text{diag}(y_u,y_c,y_t) 
\end{align}
Therefore, for the CKM matrix of quark mixing we obtain
\begin{align}\label{CKMF}
V_\mathrm{CKM} = O_u^T F O_d =  \:
O_u^T
\left(\begin{array}{ccc}
e^{i(\tilde{\beta}+\tilde{\delta})}& 0& 0 \\ 0 & e^{i \tilde{\beta}} &0 \\0&0&1
\end{array}\right)
O_d 
\end{align}
where the matrix $F=F_u^\ast F_d$ without loss of generality 
can be parameterized by the two phases $\tilde\beta$ and $\tilde\delta$ 
while the orthogonal matrices $O_{u,d}$ can be parametrized  as
\begin{align}\label{Od}
& O_d=O_{d23}O_{d13}O_{d12}= \left(\begin{array}{ccc}
 1 & 0    & 0 \\
 0 & c^d_{23} & s^d_{23} \\
 0 & -s^d_{23} & c^d_{23}
\end{array}\right) 
\left(\begin{array}{ccc}
c^d_{13} & 0 & s^d_{13} \\
0    & 1 & 0 \\
-s^{d}_{13}  & 0 & c^d_{13}
\end{array}\right)
\left(\begin{array}{ccc}
c^d_{12} &   s^d_{12} & 0 \\
 -s^d_{12}   & c^d_{12} & 0  \\
 0 & 0 &1
\end{array}\right)
\end{align}
with $c^d_{ij}=\cos\theta^d_{ij}$ and $s^d_{ij}=\sin\theta^d_{ij}$,
and analogously for up-quarks, with $O_u=O_{u23}O_{u13}O_{u12}$.
The rotations of right-handed states $O'_{u,d}$, 
can be parameterized in the same way,
with sines $s^{d(u)'}_{ij}$ and cosines $c^{d(u)'}_{ij}$.

Hence, $\widetilde{Y}_d$ contains four parameters, $A_d,B_d,C_d$ and $x_d$, 
which determine the three Yukawa eigenvalues $y_{d,s,b}$ and the three rotation angles in $O_d$. 
Analogously, the four parameters in $\widetilde{Y}_u$ determine the Yukawa eigenvalues 
$y_{u,c,t}$ and the three angles in $O_u$. 
Therefore, we have $10$ real parameters $A_{u/d},B_{u/d},C_{u/d},x_{u/d}$ 
and two phases $\tilde{\beta},\tilde{\delta}$
which 
have to match $10$ observables, the $6$ Yukawa eigenvalues   
and $4$ independent parameters of the CKM matrix \eqref{ckmSP}.

The Yukawa eigenvalues and rotation matrices $O$ and $O'$ can be found by 
considering the ``symmetric" squares respectively of the Yukawa matrices $\widetilde{Y}_f^T \widetilde{Y}_f$ and 
$\widetilde{Y}_f\widetilde{Y}_f^T$, $f=u,d$. 
In doing so, we obtain the following relations 
\begin{align}\label{secular} 
&  C^2 +  (x^2+x^{-2})B^2 + 2A^2   = Y_3^2 + Y_2^2 + Y_1^2 
  \nonumber \\
&  B^4 + 2C^2 A^2  + (x^2+x^{-2})B^2A^2 + A^4 
 = Y_3^2 Y_2^2 +  Y_3^2 Y_1^2 + Y_2^2 Y_1^2 \nonumber \\
&  
A^2 C = Y_1 Y_2 Y_3
\end{align}
where we omit the indices $f=u,d$ and imply $Y_{1,2,3}=y_{u,c,t}$ for the Yukawa 
eigenvalues of upper quarks and $Y_{1,2,3}=y_{d,s,b}$ for down quarks.

It is useful to expand the parameters
having in mind the approximate hierarchy 
$y_t : y_c : y_u\sim 1 : \epsilon_u : \epsilon_u^2$ and
$y_b : y_s : y_d\sim 1 : \epsilon_d : \epsilon_d^2$, where
it can also be noted that phenomenologically the rough relation $\epsilon_u \sim \epsilon_d^2$ applies.
In leading order approximation (up to corrections of order 
$\epsilon \sim Y_2/Y_3\sim Y_1/Y_2$) we have (see also ref. \cite{Rossi1})
\begin{align}\label{leading}
C\approx Y_3 , \quad B \approx \sqrt{Y_2Y_3}, \quad A \approx \sqrt{Y_1Y_2}
\end{align}
so that $C_f : B_f : A_f \sim 1 : \epsilon_f^{1/2} : \epsilon_f^{3/2}$. 
Since these ratios in fact reflect the hierarchy in the horizontal 
symmetry breaking \eqref{vevs}, $C: B : A \sim V_3 : V_2 :V_1$, 
this means that the inter-family mass hierarchy 
 can actually be induced by a milder hierarchy between the VEVs. 
The matrix entries $A_f$, $B_f$ and $C_f$ depend on the deformation $x_{f}$ only at higher orders in $\epsilon_{f}$:
\begin{align}
& 
 \frac{A}{Y_3} = \sqrt{\frac{Y_1 \, Y_2  }{ Y_3^2 }\frac{1  }{ c^{f\prime }_{23} c_{23}^{f} }} +\mathcal{O}(\epsilon^{7/2}) \, , \qquad 
\frac{B}{Y_3} = \sqrt{ \frac{Y_2}{Y_3} \left[1 -\frac{1}{2}\frac{Y_1}{Y_2}\left(\frac{c^{f}_{23}}{c^{ f\prime}_{23}}+\frac{c^{ f\prime}_{23}}{c^{f}_{23}}\right)  \right] } 
+\mathcal{O}(\epsilon^{7/2})
\, ,  \nonumber \\ &
 \frac{C}{Y_3} =  \sqrt{1-\frac{B^2}{Y_3^2} \left(x^2+\frac{1}{x^2}\right) + \frac{Y_2^2}{Y_3^2}-2\frac{A^2}{Y_3^2}} 
 +\mathcal{O}(\epsilon^{4})
 \,.
\end{align}

On the other hand, considering again the hierarchy $Y_3 : Y_2 : Y_1\sim 1 : \epsilon : \epsilon^2$, 
the rotation angles in \eqref{Od} appear to be small, so that $c^{d,u}_{ij}\approx 1$, 
and in the leading approximation we obtain 
\begin{align}
 s^f_{23} \approx \frac{x^{-1}B}{C} \approx \frac{1}{x}\sqrt{\frac{Y_2}{Y_3} }, \quad 
 s^f_{12} \approx \frac{AC}{B^2} \approx \sqrt{\frac{Y_1}{Y_2} } , \quad 
 \frac{s^f_{13}}{s^f_{23} }\approx  \frac{x^2 A}{C} \approx  x^2 s^f_{12}  \frac{Y_2}{Y_3}  
\end{align} 
so that $s^f_{23}, s^f_{12} \sim \epsilon_f^{1/2}$ and $s^f_{13} \sim \epsilon_f^{2}$, $f=u,d$. 
More precisely, we have 
\begin{align}
& \tan ( 2\theta_{23}^f )=\frac{2}{x}\sqrt{\frac{Y_2-Y_1}{Y_3}}\frac{\sqrt{1-(x^{-2}+x^2)\frac{Y_2-Y_1}{Y_3}+\frac{Y_2^2}{Y_3^2}}}{1-\frac{2}{x^2}\frac{Y_2-Y_1}{Y_3}+\frac{Y_2^2}{Y_3^2}}+\mathcal{O}(\epsilon^{7/2}) \nonumber \\
& \tan ( 2\theta_{12}^f)=-2\sqrt{\frac{Y_1\, c^{f\prime}_{23} }{ Y_2 \, c^f_{23} }}\, \frac{1}{1+\frac{1}{2}\frac{Y_1}{Y_3}(3 x^2-x^{-2})} +\mathcal{O}(\epsilon^{7/2}) \nonumber \\
& \tan (2\theta_{13}^f)=\frac{2A s^{f\prime}_{23}}{Y_3}+\mathcal{O}(\epsilon^{4}) =
2x \frac{Y_2}{Y_3}\sqrt{\frac{Y_1}{Y_3}}\left[1-\frac{1}{2}\frac{Y_1}{Y_2}-\frac{1}{4}\left(x^2-\frac{1}{x^2}\right)\frac{Y_2}{Y_3} \right]+\mathcal{O}(\epsilon^{4}) 
\end{align}
%
%%%%%%%%%%%%%%%%%%%%%%%%%%%%%%%%%%%%%%%%%%%%%%%%%%%%%%%%%%%
which, up to relative corrections of order $O(\epsilon^2)$, correspond to: 
\begin{align}
& s^f_{23} \approx \frac{1}{x}\sqrt{\frac{Y_2}{Y_3} } \Big( 1-\frac{1}{2} x^2 \frac{Y_2}{Y_3} -\frac{1}{2}  \frac{Y_1}{Y_2} \Big)
+\mathcal{O}(\epsilon^{5/2})
, \quad \nonumber \\
& s_{12}^f \approx -\sqrt{\frac{Y_1\, c^{f\prime}_{23} }{ Y_2 \, c^f_{23} }}
\left[1-\frac{3}{2}\frac{Y_1\, c^{f\prime}_{23} }{ Y_2 \, c^f_{23} }\right]+\mathcal{O}(\epsilon^{5/2}) , \quad \nonumber \\
& s^f_{13} \approx \frac{A s^{f\prime}_{23}}{Y_3}+\mathcal{O}(\epsilon^{4}) =
x \frac{Y_2}{Y_3}\sqrt{\frac{Y_1}{Y_3}}\Big[1-\frac{1}{2}\frac{Y_1}{Y_2}-\frac{1}{4}\big(x^2-\frac{1}{x^2}\big)\frac{Y_2}{Y_3} \Big]+\mathcal{O}(\epsilon^{4}) 
\label{sines2}
\end{align} 
The expressions for $\theta^{f\prime}_{23}$, $\theta^{f\prime}_{12}$ and $\theta^{f\prime}_{13}$ are the same with the replacement $x\rightarrow 1/x$,
$s^{f\prime}_{23}\rightarrow s^f_{23}$, $c^{f\prime}_{23}\rightarrow c^f_{23}$.

These equations show that the yukawa matrix elements $A_{u/d},B_{u/d},C_{u/d},x_{u/d}$ and 
the rotation angles in the matrices $O_{u,d}$ can be computed 
respectively in terms of the Yukawa ratios $y_s/y_b$, $y_d/y_s$ and $y_u/y_c$, $y_c/y_t$  
and the `deformation' parameters $x_d$ and $x_u$.

Then,
up to relative corrections $O(\epsilon_d^2 )$, 
we have for the CKM matrix elements
\begin{align}
& |V_{us}| = \left|\, s_{12}^d-   s_{12}^u \, c_{12}^dc_{23}^d \, e^{-i\tilde{\delta}}\, \right|
+\mathcal{O}(\epsilon^{5/2}) \, , \nonumber 
\\&
|V_{cb}| = \left|\, s_{23}^d - s_{23}^u \, c_{23}^d \,  e^{-i \tilde{\beta}}\, \right|
+\mathcal{O}(\epsilon^{5/2}) \, ,  \nonumber 
\\
& |V_{ub}| = \left| s_{13}^d\, e^{i\tilde{\delta}} -s_{12}^u \,  \left(  s_{23}^d\, c_{23}^u - s_{23}^u \, c_{23}^d \,e^{-i\tilde{\beta}}\, \right)  \right|
+\mathcal{O}(\epsilon^{4}) \, .
\label{vamanoeps}
\end{align}
It can be noticed that for fixed values of the asymmetries $x_d$, $x_u$,
$V_{us}$ depends on the phase $\tilde{\delta}$ while $V_{cb}$ only on the phase $\tilde{\beta}$.
It is also worth noting that for $x_d=1$, the contribution of $s_{13}^d$ in $|V_{ub}|$ is negligible and the Fritzsch texture implies the prediction
$|V_{ub}/V_{cb}|\approx \sqrt{y_u / y_c } $.
Similar considerations can be inferred for the other off-diagonal elements
\begin{align}
& |V_{cd}| = \left|\, s_{12}^d\, c_{23}^d -   s_{12}^u \, c_{12}^d \, e^{i\tilde{\delta}}\,+ 
s_{12}^ds_{23}^ds_{23}^u\, e^{-i\tilde{\beta}} \right|
+\mathcal{O}(\epsilon^{5/2}) \, , \nonumber 
\\&
|V_{ts}| = \left|\left( s_{23}^d  - s_{23}^u \, c_{23}^d  \,  e^{i \tilde{\beta}} \right) c_{12}^d \right|
+\mathcal{O}(\epsilon^{5/2}) \, ,  \nonumber 
\\
& |V_{td}| = \left| s_{13}^d\, e^{i\tilde{\delta}} -s_{12}^d \, 
 \left( s_{23}^d  - s_{23}^u \, c_{23}^d  \,  e^{i \tilde{\beta}} \right)   \right|
+\mathcal{O}(\epsilon^{3}) \, .
\label{vamanoepsb}
\end{align}
with the prediction $|V_{td}/V_{ts}|\approx \sqrt{y_d / y_s } $ for $x_d=1$.
As regards the complex part of $V_\text{CKM}$, 
we can consider the rephasing-invariant quantity
$J=-\text{Im}(V^*_{us}V^*_{cb}V_{ub}V_{cs}) $,
the Jarlskog invariant.
In our scenario 
we have
\begin{align}
 J= & -\sin\tilde{\delta} \, s_{12}^u s_{12}^d \, \left[  (s_{23}^d)^{2}\, c_{23}^d c_{12}^d 
- 2 \cos\tilde{\beta} \, s_{23}^d s_{23}^u + (s_{23}^u)^{2}  \right] +  \nonumber  \\
&\, +  \left( \sin\tilde{\delta} s_{12}^u s_{23}^d+ \sin\tilde{\beta} s_{12}^d s_{23}^u \right) s_{13}^d+\mathcal{O}(\epsilon_d^4)
\label{Jamano}
\end{align}

\section{Analysis of the Yukawa parameters and CKM mixing}
\label{analysis}

\subsection{Observables}
\label{observables}

\begin{table}[t]
\centering
\begin{tabular}{| l l l  |}
\hline
 $m_s/m_{ud}$ 
                & $ 27.31(10) $ & PDG \cite{PDG22}$^*$ \\
$m_u/m_{d}$             
                      & $ 0.477(19)$ &   PDG \cite{PDG22}$^*$ \\
$m_c/m_{s}$ & $ 11.768(34)$ & FLAG $N_f=2+1+1$ \cite{FLAG21} \\ 
$m_b/m_{s}$ & $ 53.94(12)$ & Bazavov et al. 2018 \cite{Bazavov18} \\ 
$m_b/m_{c}$ & $ 4.579(9)$ & PDG \cite{PDG22}\\
$Q$ & $ 22.9(4)$ & PDG \cite{PDG22}$^*$  \\
 & $ 22.1(7)$ & phenomenological \cite{Colangelo} \\
 $M_t$ & $ 172.69\pm 0.30 $ GeV & PDG \cite{PDG22}\\
$m_b(m_b)$ & $4.203(11)$ GeV & FLAG 2021  $N_f=2+1+1$ \cite{FLAG21} \\
$M_Z$ & $91.1876(21)$ GeV   & PDG \cite{PDG22}\\
$\alpha_s(M_Z)$ & $ 0.1185(16) $  & PDG \cite{PDG22}\\
$\alpha(M_Z)^{-1}$ & $ 127.951(9) $  & PDG \cite{PDG22}\\
$\sin^2\theta_W(M_Z)$ & $ 0.2299(43) $  & PDG \cite{PDG22}\\
\hline
\end{tabular}
\caption{\label{masses}  
Determinations of quark mass ratios used in this work. 
In the first line, $m_{ud}=(m_u+m_d)/2$. \\
$^*$ Value adopted by Particle Data Group (PDG), averaging $N_f=2+1+1$ and $N_f=2+1+1$ flavours lattice results \cite{FLAG21}.
}
\end{table}

\begin{table}[t]
\centering
\begin{tabular}{| lr  @{\hspace{4\tabcolsep}}  l r |}
\hline
Quantity & Value & Quantity & Value \\
\hline
$|V_{ud}|$ & $0.97373(31)$  & $|V_{cs}|$ & $0.975(6)$     \\
$|V_{us}|$ & $0.2243(8)$  &$|V_{cd}|$ & $0.221(4)$   \\
$|V_{ub}|$ & $0.00382(20)$  & $|V_{td}|$ & $0.0086(2)$  \\
$|V_{cb}|$ & $0.0408(14)$  & $|V_{ts}|$ & $0.0415(9)$  \\
$|V_{td}/V_{ts}|$ & $0.207(3)$   & $|V_{ub}/V_{cb}|$ & $0.084(7)$  \\
\hline
\end{tabular}
\caption{\label{values}  
Magnitudes and phases of CKM elements as quoted by Particle Data Group \cite{PDG22}.
}
\end{table}
\begin{table}[t]
\centering
\begin{tabular}{| lr   |}
\hline
Parameter & Global fit value \cite{PDG22} \\
\hline
$\sin\theta_{12}$ & $0.22500\pm 0.00067$ \\
$\sin\theta_{23}$ & $0.04182^{+0.00085}_{-0.00074}$ \\
$\sin\theta_{13}$ & $0.00369\pm 0.00011$ \\
$J$ & $(3.08^{+0.15}_{-0.13})\times 10^{-5}$ \\
$\delta$ & $1.144\pm 0.027$ \\
\hline
\end{tabular}
\caption{\label{global}  
Result of global fit for CKM parameters, including constraints implied by the unitarity of the three generation CKM matrix,
as reported by Particle Data Group \cite{PDG22}.
}
\end{table}

\begin{figure}
\centering
\includegraphics[width=0.7\textwidth]{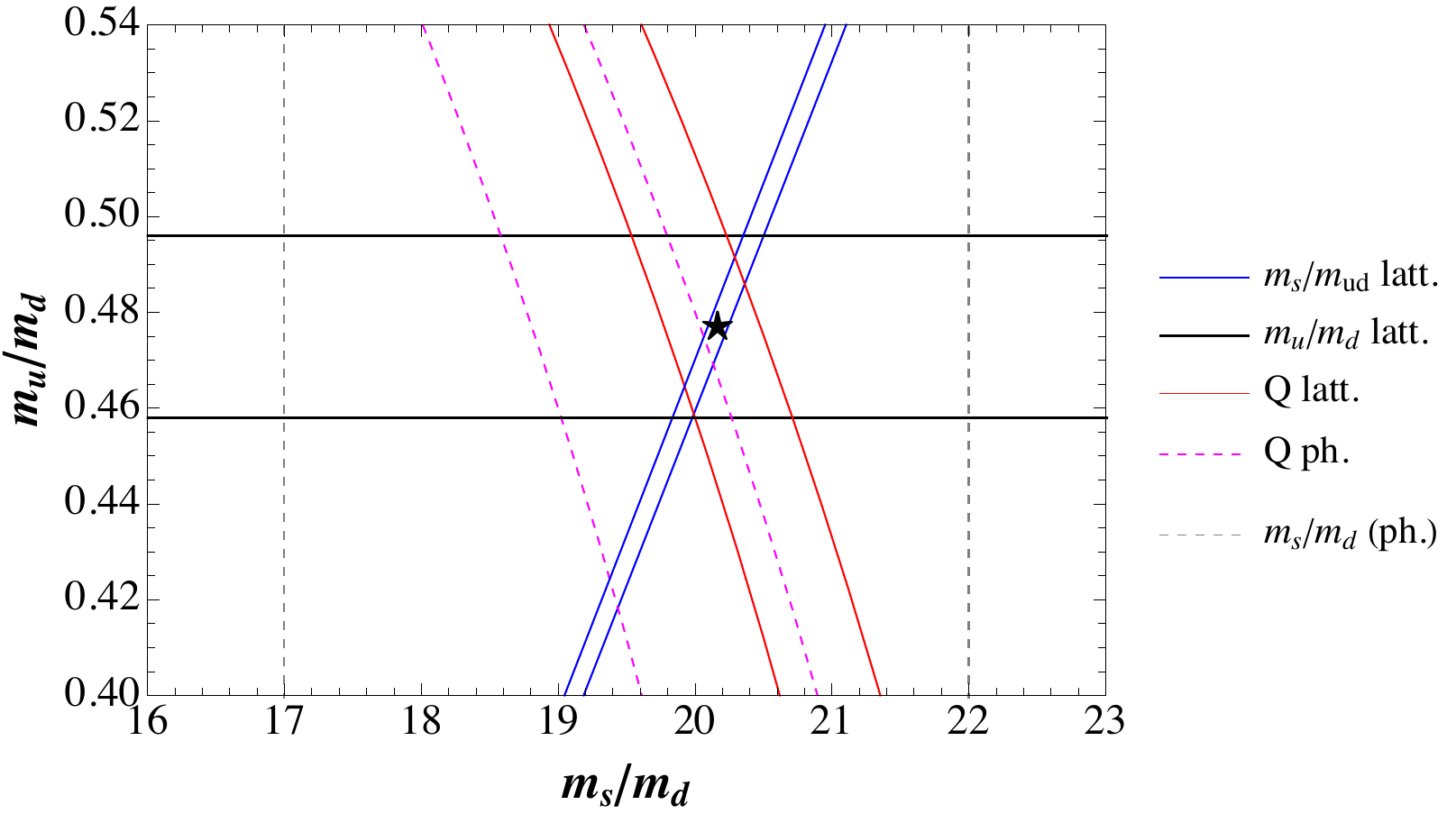}
\caption{\label{mumd} Light quarks mass ratios (see also table \ref{masses}).
Black continuous lines show the average of three and four flavours lattice determinations 
of the ratio $m_u/m_d$; 
blue lines represent the average of three and four flavours lattice determinations of the ratio $m_s/m_{ud}$, $m_{ud}=(m_u+m_d)/2$; red lines are obtained from the relation 
$Q^2=(m_s^2-m_{ud}^2)/(m_d^2-m_u^2)$, using lattice determinations of quark mass 
ratios. 
We also indicate the phenomenological determinations $Q= 22.1(7)$ \cite{Colangelo} (dashed magenta)
and $m_s/m_d = 17$--$22$ \cite{PDG22} (dashed grey).
The black star represents the central value $(m_u/m_d,m_s/m_d )=(0.477, 20.17)$.
}
\end{figure}

The input values in our analysis will be the ratios of the Yukawa eigenvalues and 
the CKM matrix elements. 
More specifically, since we do not need to make assumptions on the energy scale at which the Yukawa matrices assume the Fritzsch form, we want to reproduce Yukawas ratios and CKM elements at different energy scales.
Yukawa matrices evolve according to the renormalization group equations, as a function of the energy scale.
For energy scales $\mu \lesssim m_t$, 
the running is basically determined by the strong coupling $\alpha_s(\mu)$ 
and QCD renormalization factors cancel in quark-mass ratios. 
We can derive the ratios of Yukawa couplings
through the ratios of  running quark masses at $\mu = m_t$. 
The latter ratios 
can be deduced from the data collected in table \ref{masses}. 
For up quarks we also need the ratio $m_t/m_b$,
from renormalization group equations (see for example ref. \cite{Chetyrkin}) 
we obtain $m_t/m_b= 59.46\pm 0.55$. 
Then, 
at $\mu=m_t$ we have
\begin{align}
& \frac{m_d}{m_s} = \frac{1}{20.17\pm 0.27} \, , && \frac{m_u}{m_c}=\frac{1}{498 \pm 21}   \nonumber \\
&\frac{m_s}{m_b} = \frac{1}{53.94\pm 0.12}   \, , && \frac{m_c}{m_t}=\frac{1}{272.3 \pm 2.6} \, , 
\label{input}
\end{align}
where we extracted the value  $m_s/m_d$ through the relation $m_s/m_d=m_s/m_{ud} \, (m_u/m_d + 1)/2$ 
(blue band in figure \ref{mumd}).
The 
main source of uncertainty in the mass ratios belongs to 
the ratio $r_{ud}=m_u/m_d$, which affects the ratios $m_d/m_s (r_{ud}) $, $m_u/m_c (r_{ud}) $.\footnote{
In the following we are going to neglect the other small errors contributing in eq. \eqref{input} 
and consider only this larger uncertainty.}

For energy scales 
$\mu$ larger than $ m_t$,
the set of coupled differential equations for the running of the Yukawa and gauge couplings should be considered
(see refs. \cite{Machacek0,Machacek1,Luo}). 
Namely, the renormalization group evolution of $y_t$ at one loop reads:
\begin{align}
& \mu \frac{d y_t}{d \mu}\approx y_t  \frac{1}{16\pi^2}  \left(\frac{9}{2} y_t^2 -\frac{17}{20}g_1^2-\frac{9}{4}g_2^2 -8g_s^2 \right)
\end{align}
where $g_s$, $g_2$, and $g_1$ are normalized as in $SU(5)$, so that the electroweak gauge coupling constants are
$g_2^2/(4\pi)=g^2/(4\pi)=\alpha/\sin^2\theta_W$ and $ g_1^2/(4\pi)=\frac53 g^{\prime 2}/(4\pi)=\frac53\alpha/\sin^2\theta_W$.
Since the light generations evolve in the same way with gauge couplings and trace of Yukawa matrices,
the ratios $y_d/y_s$, $y_u/y_c$ remain invariant.
The third generation instead receives additional Yukawa contributions.
Consequently the ratios with the heaviest generations evolve as:
\begin{align}
& \mu \frac{d}{d \mu} \frac{y_c}{y_t}\approx - \frac{y_c}{y_t} \, \frac{1}{16\pi^2}   \frac{3}{2} \, y_t^2 \, ,
\qquad \mu \frac{d}{d \mu} \frac{y_s}{y_b}\approx \frac{y_s}{y_b}  \, \frac{1}{16\pi^2} \frac{3}{2} \,  y_t^2
\end{align}

As regards the CKM matrix, the mixing angles involving the third generation change according to renormalization group equations  \cite{Babu,Barger}:
\begin{align}
& \mu \frac{d V_{cb}}{d \mu}  \approx V_{cb} \,\frac{1}{16\pi^2}   \frac{3}{2} \, y_t^2 \, , \qquad
 \mu \frac{d V_{ub}}{d \mu} \approx V_{ub} \, \frac{1}{16\pi^2}  \frac{3}{2} \,  y_t^2
\label{Vren}
\end{align}
and similarly for $V_{td}$ and $V_{ts}$,
while the mixings between the first two families ($V_{us}$, $V_{ud}$, $V_{cd}$, $V_{cs}$) remain unchanged.
As regards the CP violating Jarlskog invariant, the scaling of $J$ at leading order is the same as $|V_{cb}|^2$,
$|V_{td}|^2$, etc., see eq. (\ref{Vren}).

In order for the Yukawa matrices (\ref{YFr}) to be a viable texture,
we must verify that we can obtain the correct determinations of the quark masses 
and of moduli and phases of the elements of the CKM matrix $V_\text{CKM}$.
However, 
since $V_\text{CKM}$ is unitary, there are only $4$ independent observables.
In the standard parameterization \eqref{ckmSP} these quantities correspond to:
\begin{align}
& s_{13}^2=|V_{ub}|^2 \, , \quad s_{12}^2=\frac{|V_{us}|^2}{1-|V_{ub}|^2}
\, , \quad s_{23}^2=\frac{|V_{cb}|^2}{1-|V_{ub}|^2} 
\nonumber \\ & 
J=-\text{Im}(V_{ub}V_{cs}V^*_{us}V^*_{cb})   
\label{angrephinv}
\end{align}
where we indicated the invariant $J$ instead of the 
phase $\delta$.
The results of the latest global fit are \cite{PDG22}:
\begin{align}
& s_{13}=0.00369(11) \, , \quad s_{12}=0.22500(67)
\, , \quad s_{23}=0.04182^{+0.00085}_{-0.00074}
\label{angfit}
\end{align}
and concerning CP violation:
\begin{align}
& J=(3.08^{+0.15}_{-0.13})\times 10^{-5}  
\label{Jgfit}
\end{align}
or $\delta =1.144(27)$. These values produce the CKM matrix \cite{PDG22}
\begin{align}
&\left| V_\text{CKM}\right| =\left(
\begin{array}{ccc}
0.97435\pm 0.00016 & 0.22500\pm 0.00067 & 0.00369\pm 0.00011 \\
0.22486\pm 0.00067 & 0.97349\pm 0.00016 & 0.04182^{+0.00085}_{-0.00074} \\
0.00857^{+0.00020}_{-0.00018} & 0.04110^{+0.00083}_{-0.00072} & 0.999118^{+0.000031}_{-0.000036}
\end{array}
\right)
\label{globalckm}
\end{align}
which may be compared to the determinations collected in table \ref{values}.

\subsection{Analysis and results}
\label{results}

In this section we are going to verify that asymmetric Fritzsch textures 
can predict quark masses together with moduli and phases of the mixing elements,
given present precision of 
experimental data and recent results from lattice computations. 
We will test the validity of this flavour pattern considering its formation at different energy scales.
As already noted, we have $10$ real parameters $A_{d,u}$, $B_{d,u}$, $C_{d,u}$, $x_{d,u}$, $\tilde{\beta}$, $\tilde{\delta}$ 
which have to match $10$ observables, 
the $6$ Yukawa eigenvalues and the $4$ independent parameters of the CKM matrix.

\subsubsection{Symmetric Fritzsch texture: why it does not work}
The canonical Fritzsch texture 
with $x_{u,d}=1$, employing $8$ parameters to determine $10$ observables, would be the most predictive structure.
However, it is in contradiction with the experimental data, as can be seen in refs. \cite{Kang,Ramond:1993kv}.
Before proceeding with the modified texture, we illustrate %in our formalism 
the reasons behind the %wrong predictions 
failure of the Fritzsch matrices, taking into account present precision data and using our formalism, 
in order to demonstrate how these predictions can be adjusted by a deformation of the texture.

In this scenario,
the $6$ yukawa eigenvalues and the element $V_{us}$ (or equivalently the Cabibbo angle $s_{12}$) 
can be reproduced by the symmetric Yukawa matrices by using $7$ parameters,
the $6$ yukawa elements and the phase $\tilde{\delta}$.
In fact, as it is apparent from eq. \eqref{vamanoeps}, 
besides the ratios of Yukawas, the value of $V_{us}$ 
is selected only by the phase $\tilde{\delta}$ while it 
has almost no dependence on the phase $\tilde{\beta}$.
On the contrary, the value of $V_{cb}$ is determined by the phase $\tilde{\beta}$, independently of the phase $\tilde{\delta}$. 
Along with that, in the symmetric case, $V_{ub}$ only presents a mild dependence on $\tilde{\delta}$, given the smallness
of $s_{13}^d$. Therefore, the symmetric Fritzsch texture also implies a clean prediction of the ratio 
$|V_{ub}/V_{cb}|\approx \sqrt{m_u / m_c } $, basically independent of $\tilde{\delta}$ but also of $\tilde{\beta}$. 
However, 
the experimental determinations of $|V_{cb}|$ and of the ratio $|V_{ub}/V_{cb}|$
cannot be accommodated by any of the values of the phase $\tilde{\beta}$,
regardless of the energy scale at which the Yukawa matrices assume the Fritzsch texture.
In a similar way, the predicted values of $|V_{td}|$, $|V_{ts}|$ and $|V_{td}|/|V_{ts}|$ appear to be too large.

\begin{figure}
\centering
  \begin{subfigure}{0.48\textwidth}
\includegraphics[width=\textwidth]{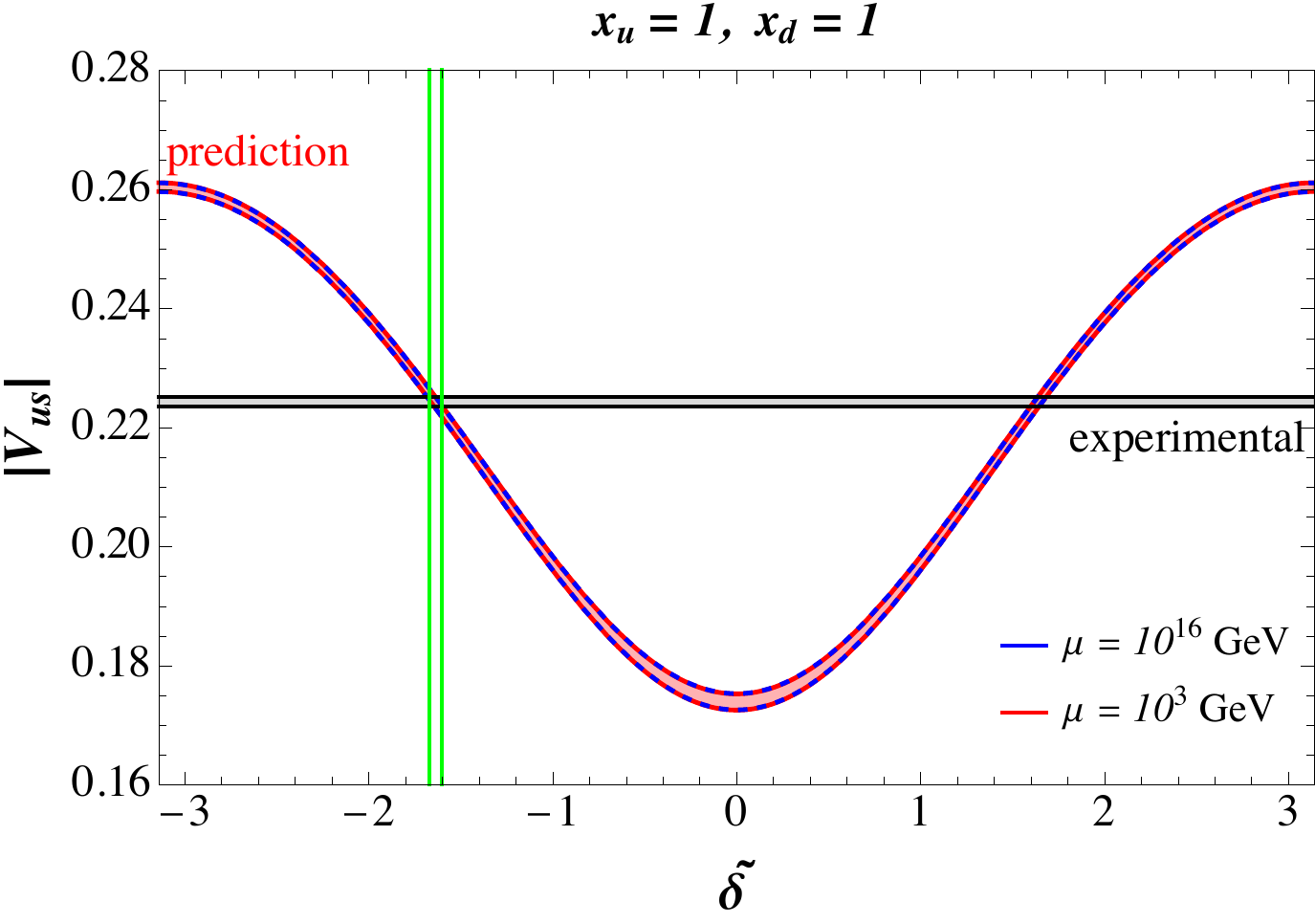}
\caption{\label{symm1}}
 \end{subfigure}
 \hfill
 \hspace{10pt}
\begin{subfigure}{0.47\textwidth}
\includegraphics[width=\textwidth]{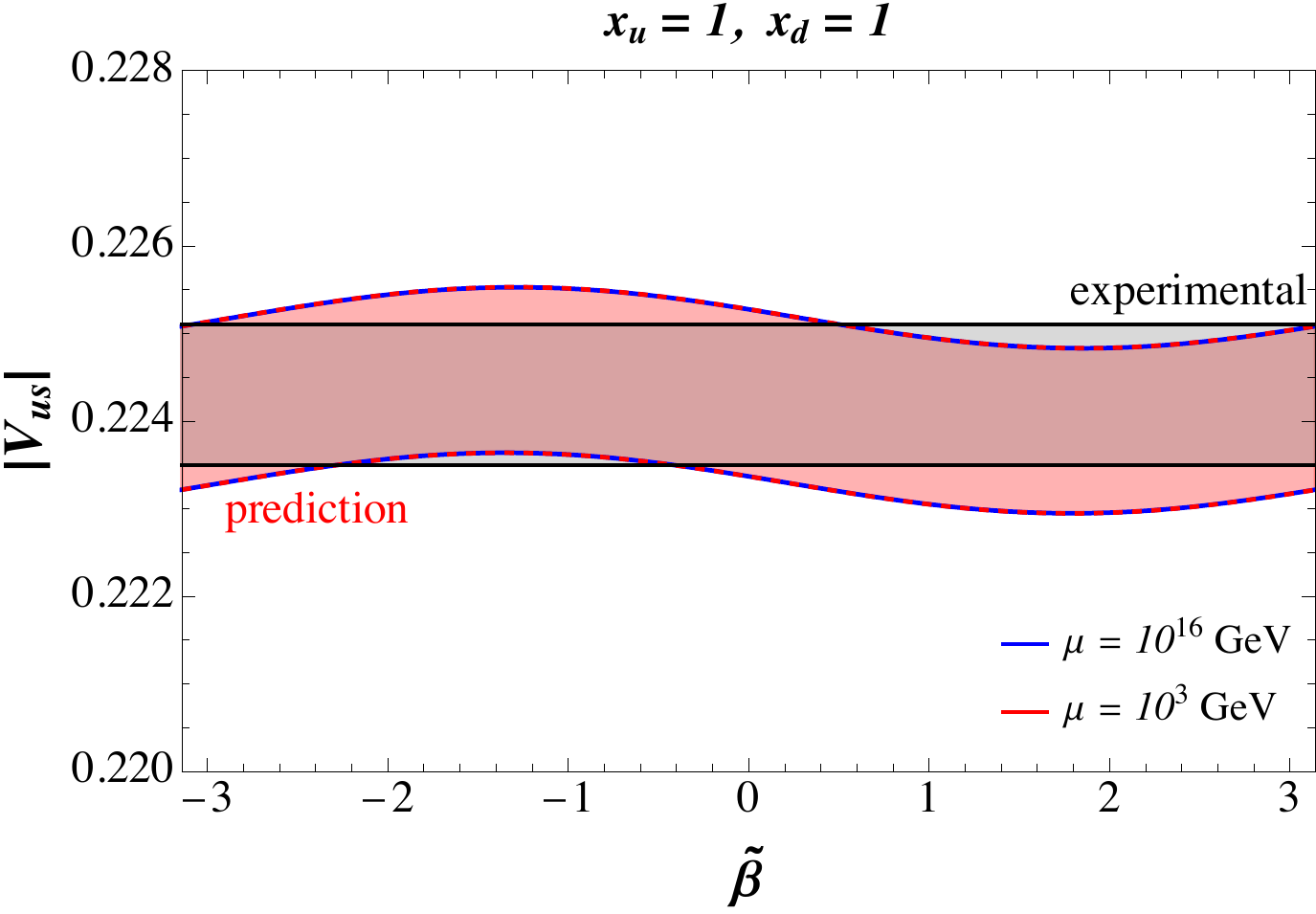}
\caption{\label{symm2}}
 \end{subfigure}
 \hfill
 \\ \vspace{10pt}
\begin{subfigure}{0.48\textwidth}
\includegraphics[width=\textwidth]{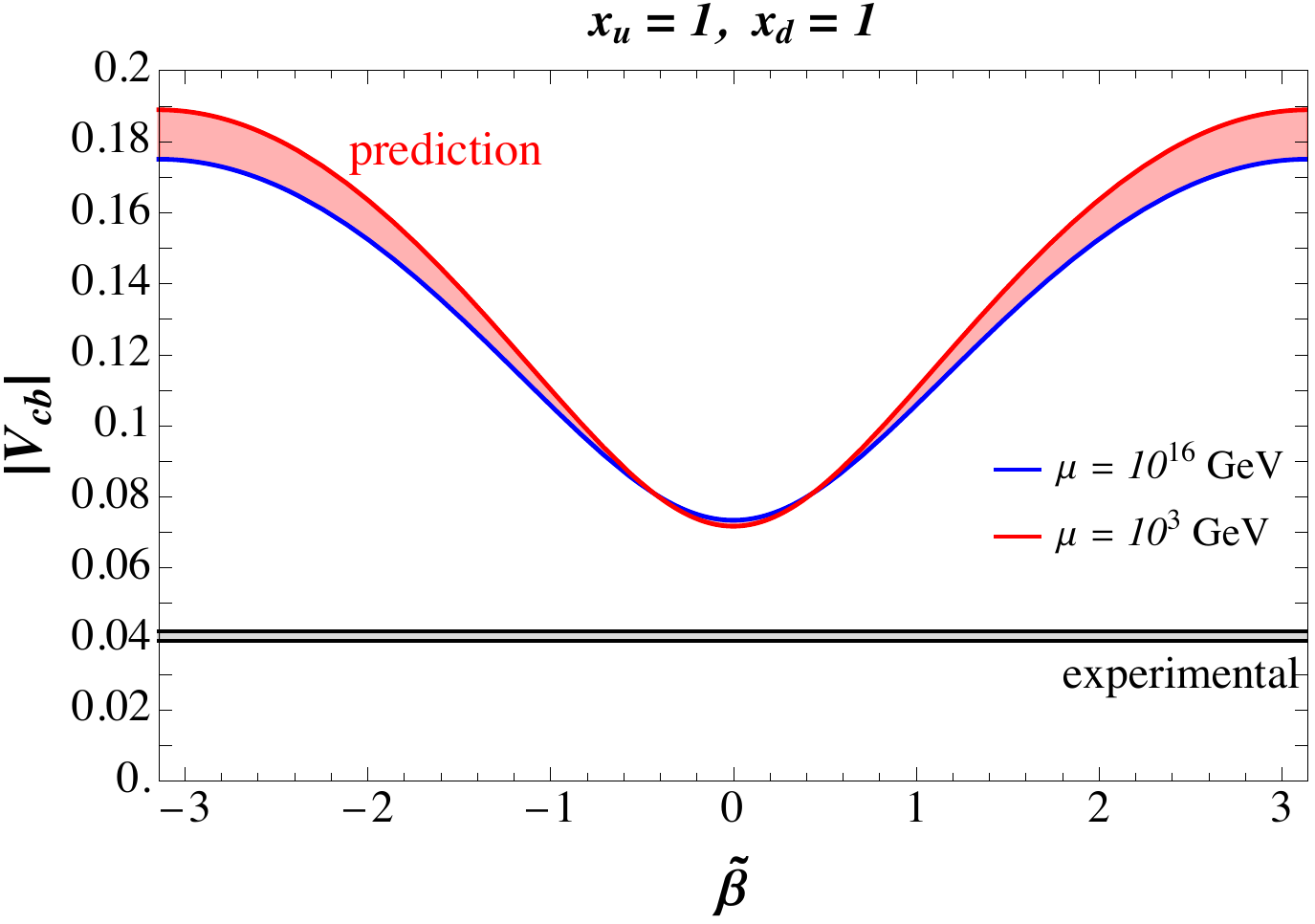}
\caption{\label{symm3}}
 \end{subfigure}
 \hfill
 \hspace{10pt}
 \begin{subfigure}{0.48\textwidth}
\includegraphics[width=\textwidth]{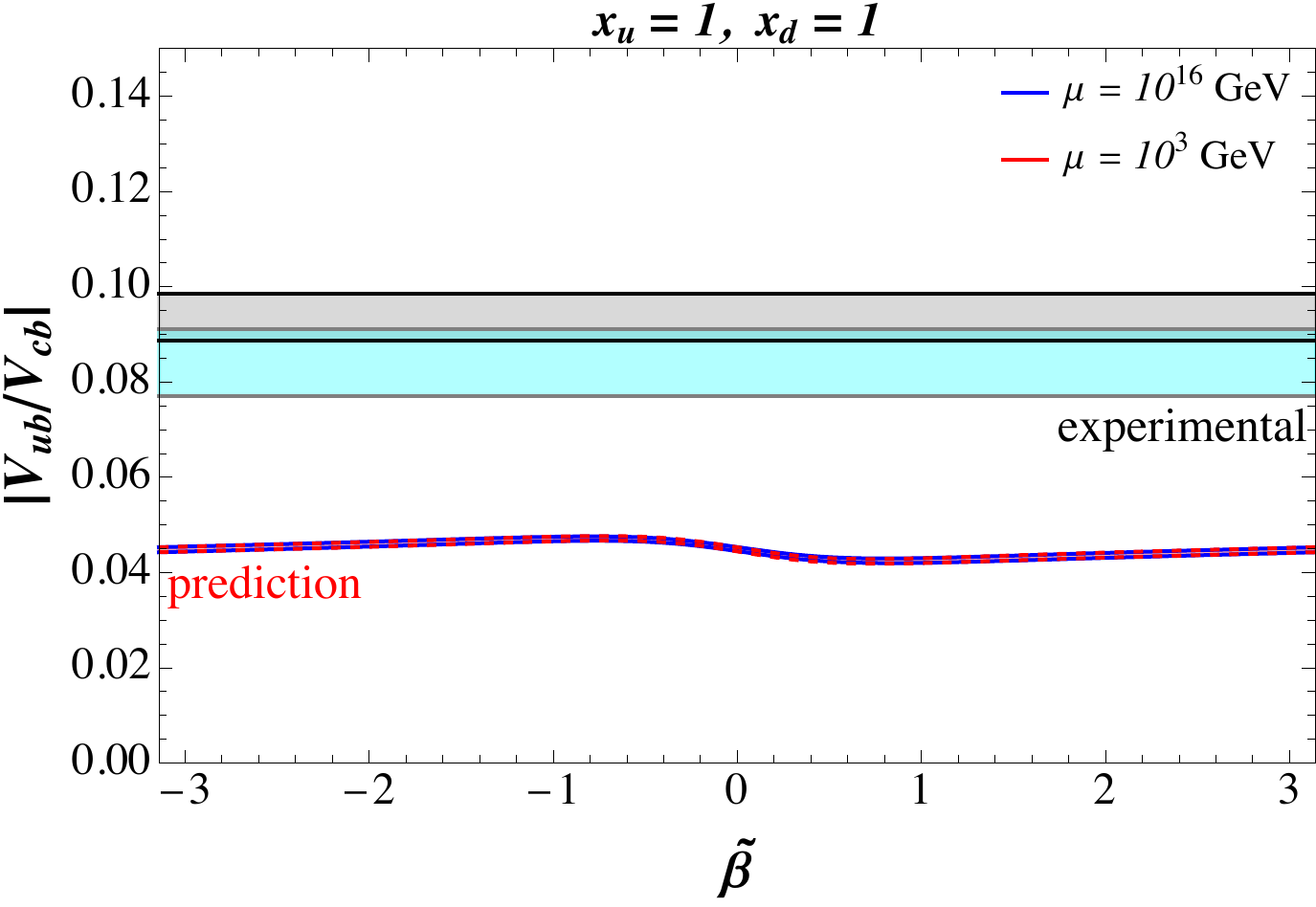}
\caption{\label{symm4}}
 \end{subfigure}
\caption{\label{symm}
Symmetric Fritzsch textures for quark Yukawa matrices do not predict the correct value of $V_{cb}$ and of
the ratio $V_{ub}/V_{cb}$. 
}
\end{figure}

We illustrate this result in figure \ref{symm}.
The value of $V_{us}$ determines the phase $\tilde{\delta}$ as shown in figure \ref{symm1}, \ref{symm2}.
In figures \ref{symm3} and \ref{symm4} we show the value of 
$V_{cb}$ and $V_{ub}/V_{cb}$ in this case, given 
$V_{us}$ within $1\sigma$ of the experimental constraint.
We indicate the predictions (red bands) assuming that the Yukawa matrices 
present the symmetric Fritzsch texture at a scale between $10^3$~GeV (blue lines) and $10^{16}$~GeV (red lines),
confronted with the experimental determinations (grey bands) at $1\sigma$ confidence level.
In figure \ref{symm4}, 
in addition to the experimental determination 
$|V_{ub}/V_{cb}|=0.094(5)$ (grey) obtained from the separate determinations of 
$V_{ub}$ and $V_{cb}$ (see table \ref{values}),
we also indicate the independent measurement of the ratio $|V_{ub}/V_{cb}|=0.084(7)$ (cyan).
The width of the prediction in this case is given by the uncertainty on the ratio $m_u/m_d$ and on $V_{us}$.
It is clear that 
the expectations implied by this scenario 
largely disagree with the experimental requirements.

We are therefore going to consider the minimally modified Fritzsch texture with asymmetry parameters $x_d$ and $x_u$.
We will pay a special attention to the $9$ parameters case with
$x_u=1$, $x_d\neq 1$ which can be motivated in the context of $SU(5)$ grand unification
and demonstrate that such a predictive ansatz can perfectly work.

\subsubsection{Asymmetric Fritzsch texture: how it works}

\begin{figure}
\centering
\begin{subfigure}{0.48\textwidth}
\includegraphics[width=\textwidth]{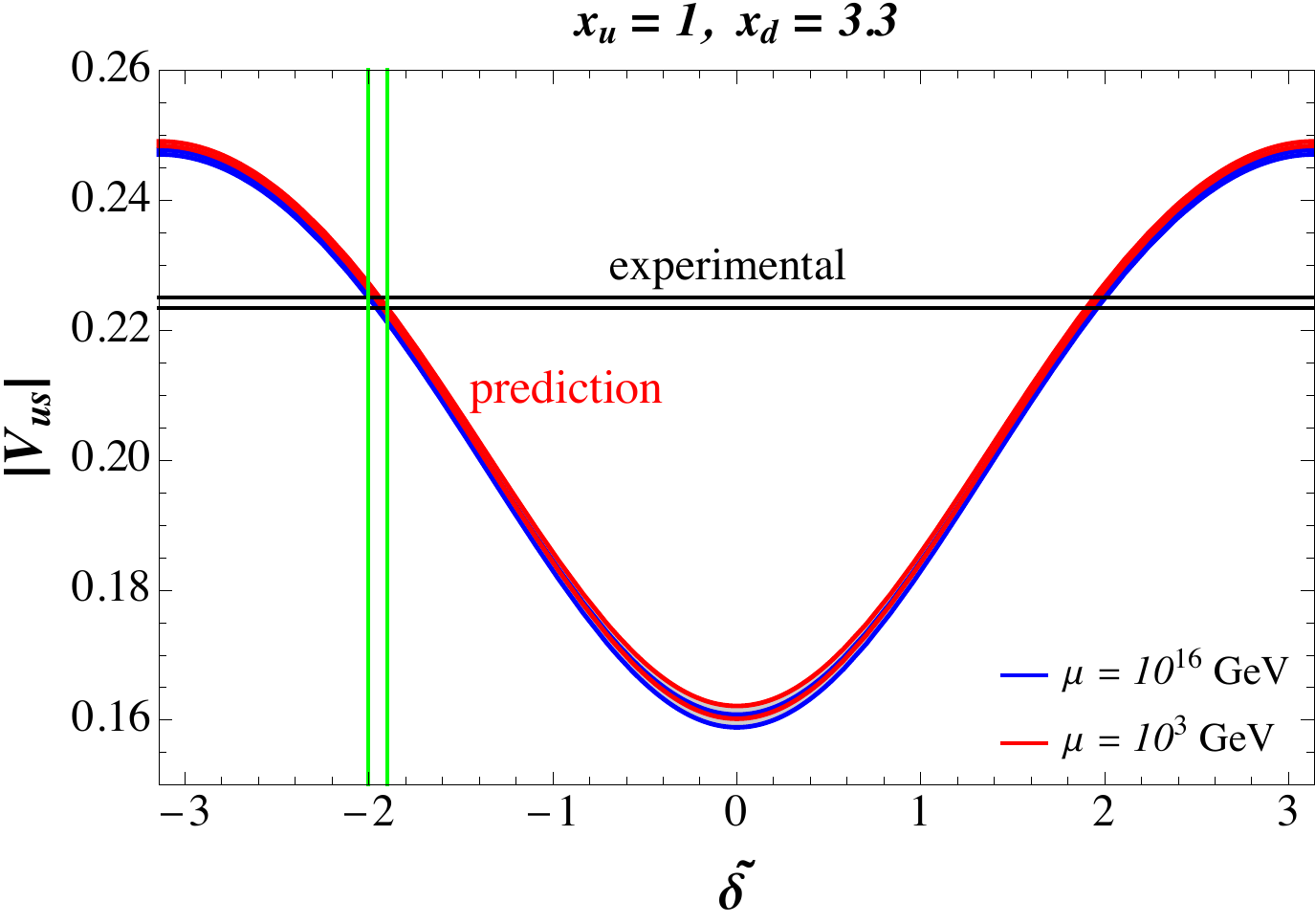}
\caption{\label{asymm1}}
 \end{subfigure}
 \hfill
\hspace{10pt}
\begin{subfigure}{0.48\textwidth}
\includegraphics[width=\textwidth]{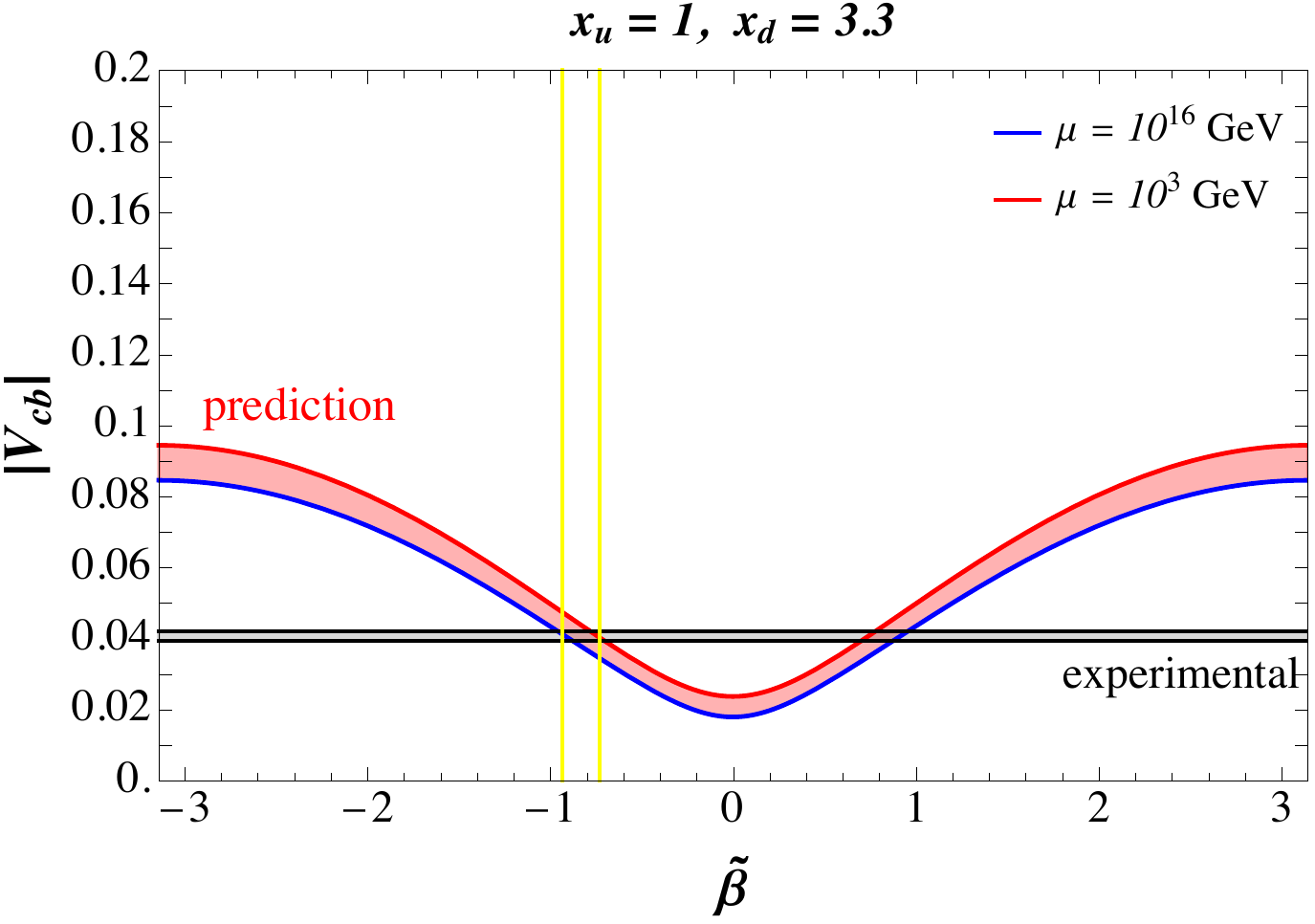}
\caption{\label{asymm2}}
 \end{subfigure}
 \hfill
 \\ \vspace{10pt}
 \begin{subfigure}{0.48\textwidth}
\includegraphics[width=\textwidth]{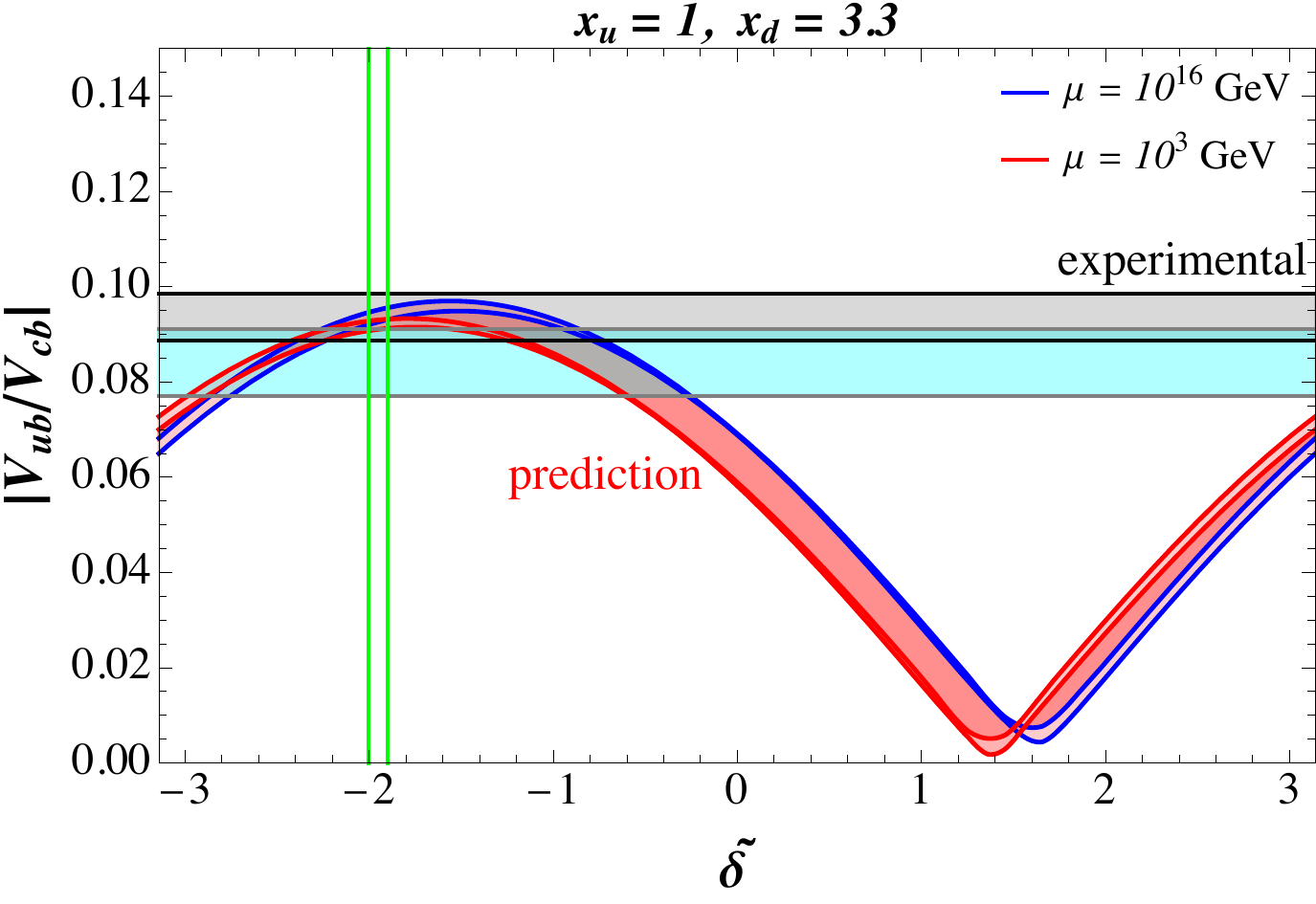}
\caption{\label{asymm3}}
 \end{subfigure}
\hfill
\begin{subfigure}{0.48\textwidth}
\includegraphics[width=\textwidth]{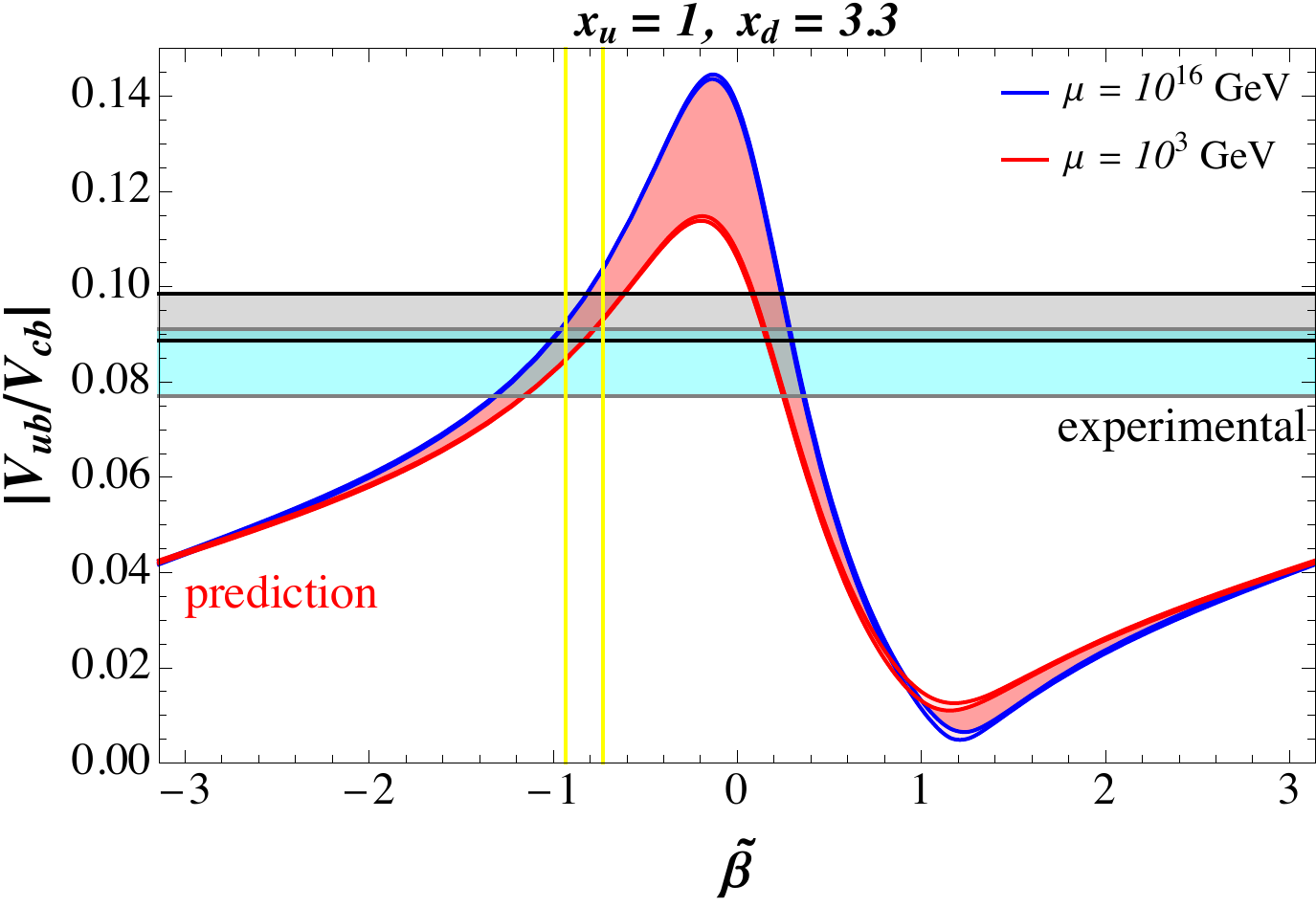}
\caption{\label{asymm4}}
 \end{subfigure}
   \hfill
 \\ \vspace{10pt}
\begin{subfigure}{0.48\textwidth}
\includegraphics[width=\textwidth]{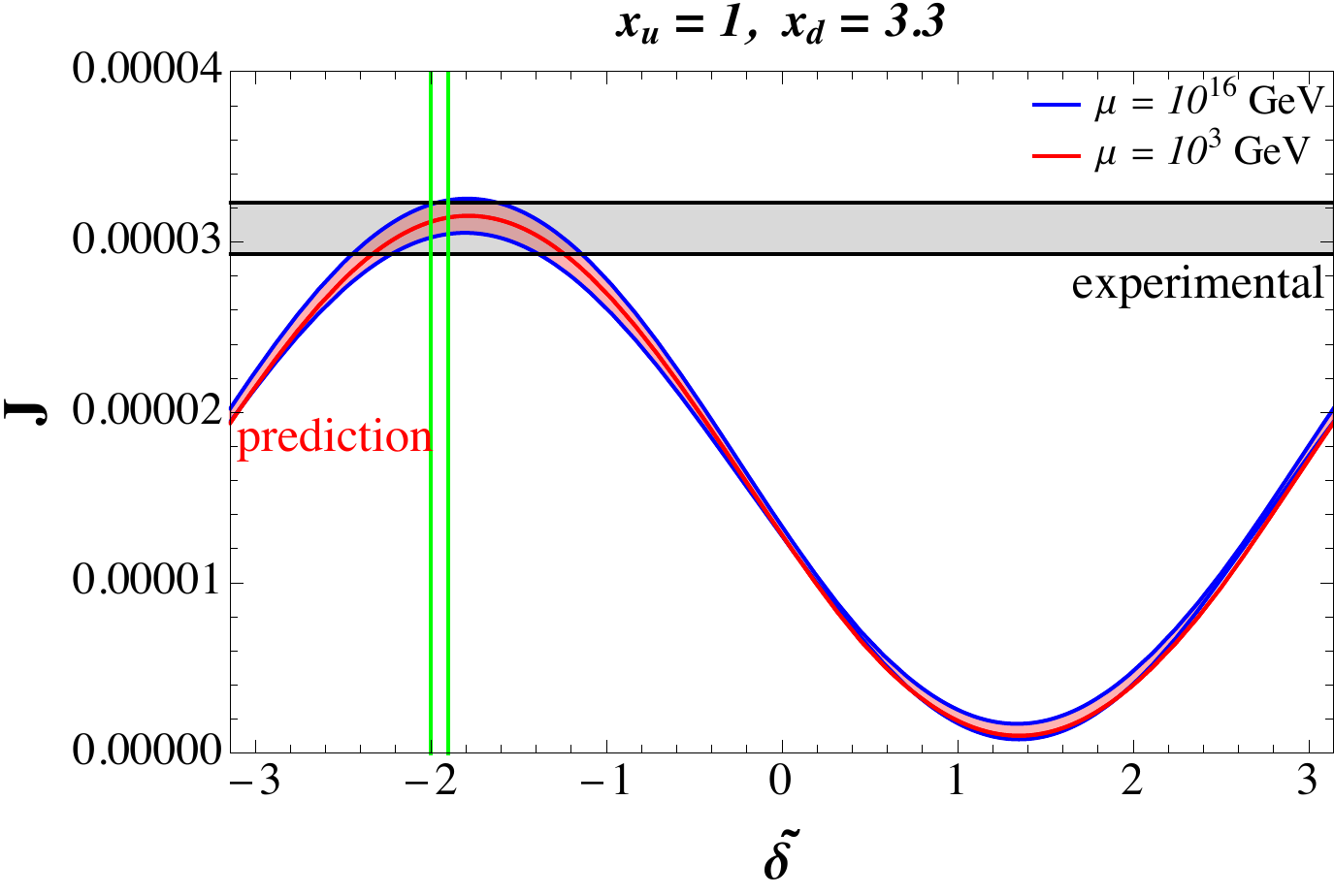}
\caption{\label{asymm5}}
 \end{subfigure}
 \hfill
\hspace{10pt}
\begin{subfigure}{0.48\textwidth}
\includegraphics[width=\textwidth]{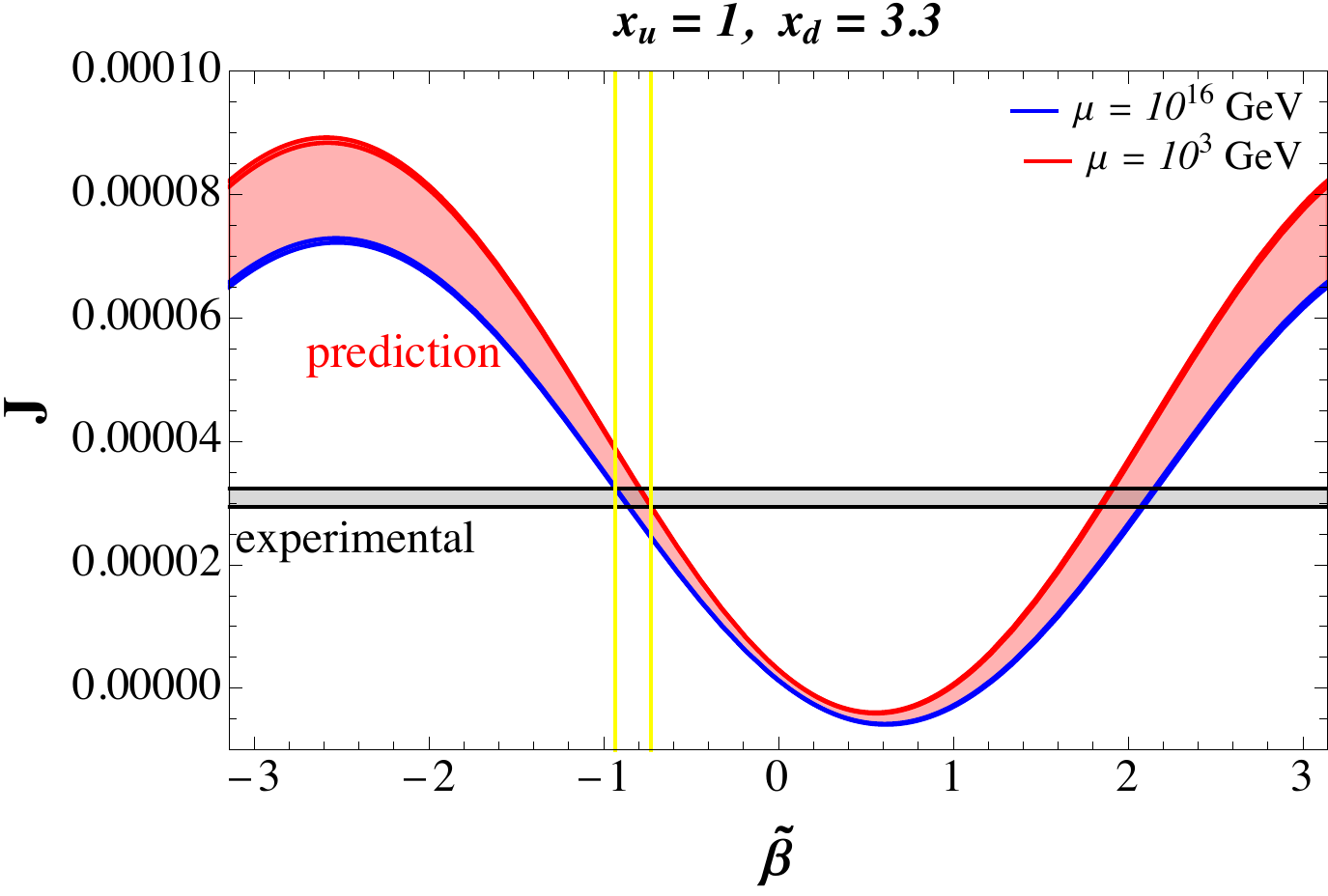}
\caption{\label{asymm6}}
 \end{subfigure}
\caption{\label{asymm}
Predictions of the asymmetric Fritzsch-like textures (see eq. \eqref{YFr}) with $x_d=3.3$, $x_u=1$,
confronted with experimental data.
}
\end{figure}

\begin{figure}
\centering
\begin{subfigure}{0.4\textwidth}
\includegraphics[width=\textwidth]{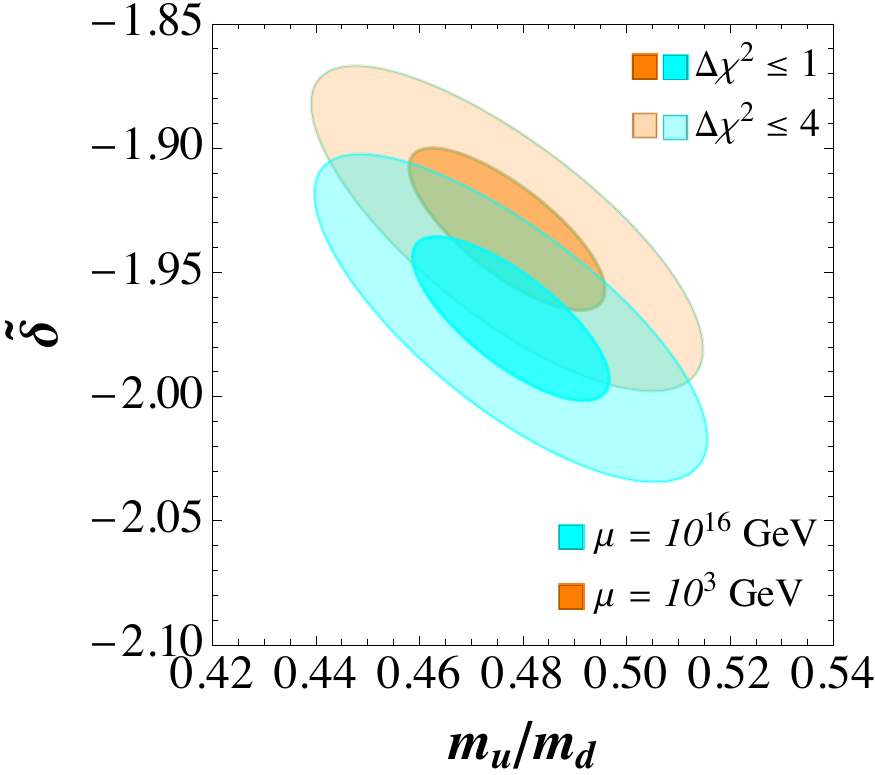}
\caption{\label{fig-delta}}
 \end{subfigure}
% \hfill
\hspace{30pt}
\begin{subfigure}{0.4\textwidth}
\includegraphics[width=\textwidth]{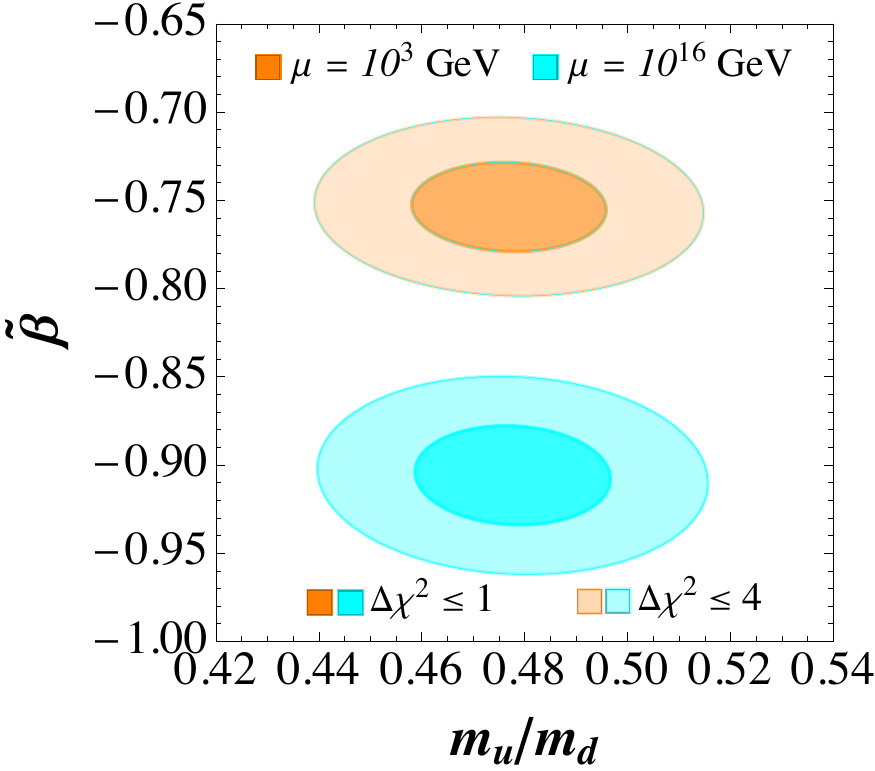}
\caption{\label{fig-beta}}
 \end{subfigure}
\caption{\label{fig-phase}
$1\sigma$ and $2\sigma$ confidence intervals of the parameters.
}
\end{figure}

\begin{figure}
\centering
\begin{subfigure}{0.48\textwidth}
\includegraphics[width=\textwidth]{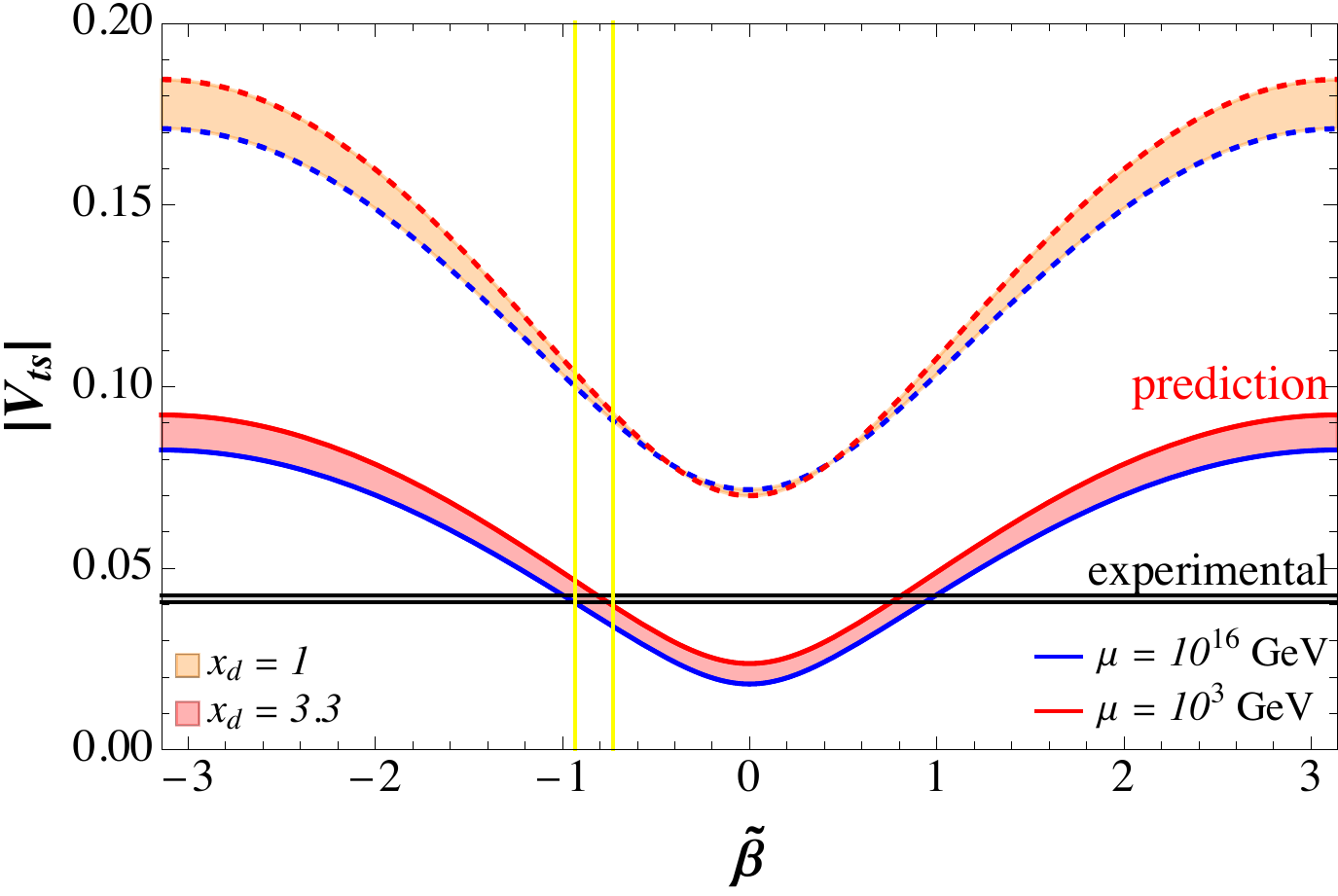}
\caption{\label{fig-vts}}
 \end{subfigure}
 \hfill
\hspace{10pt}
\begin{subfigure}{0.48\textwidth}
\includegraphics[width=\textwidth]{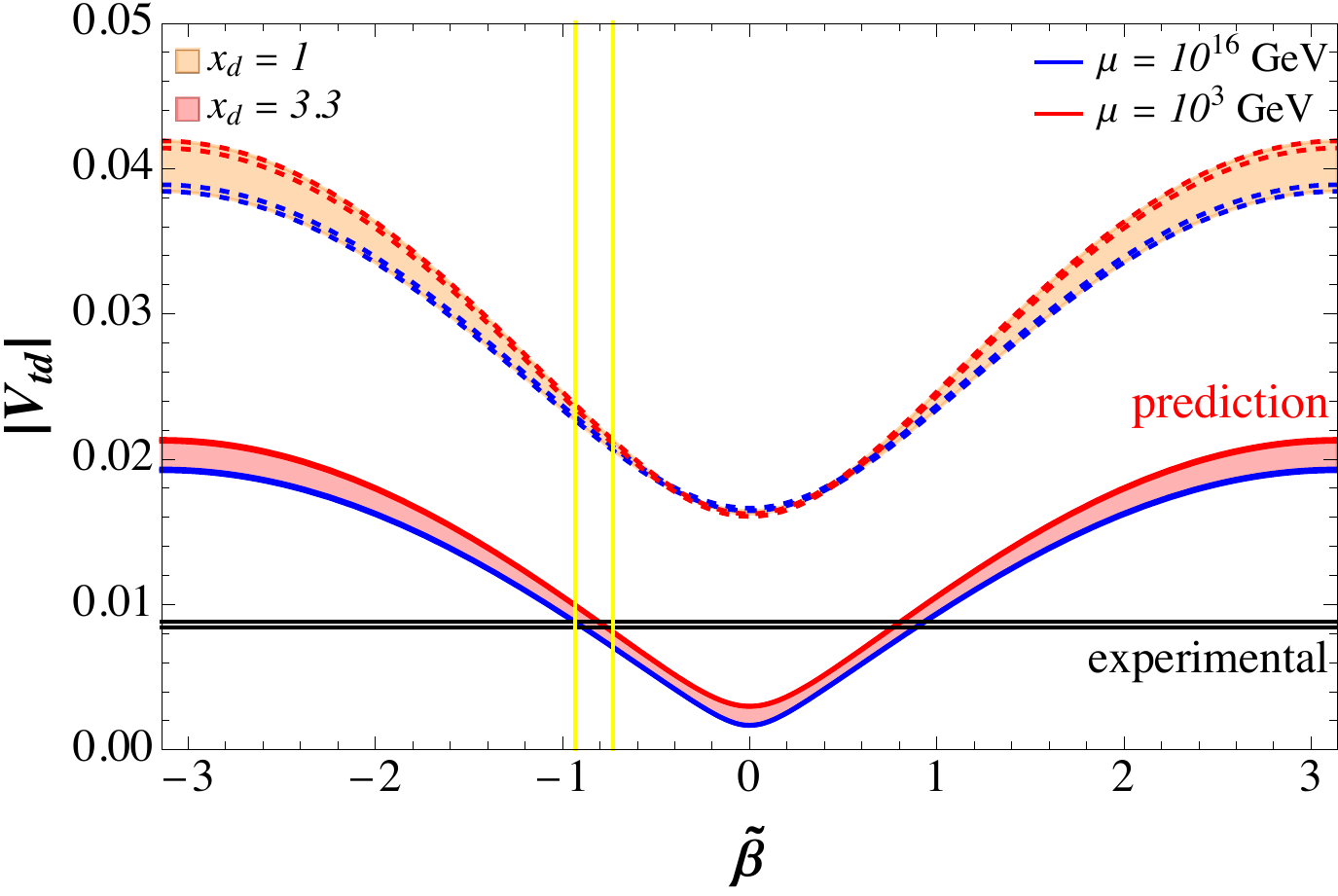}
\caption{\label{fig-vtd}}
 \end{subfigure}
\caption{\label{fig-vtdvts}
Predictions of $V_{ts}$ and $V_{td}$ from asymmetric Fritzsch textures, with paramters determined as in figure \ref{asymm}.
We also show the wrong predictions of the symmetric Fritzsch texture (orange region) 
at a scale between $10^3$~GeV (blue dashed lines) and $10^{16}$~GeV (red dashed lines).
}
\end{figure}

Let us first provide an example in order to 
convey a comparison with the expectations of the standard Fritzsch texture displayed in figure \ref{symm}.
For this purpose, we fix $x_d=3.3$ (keeping $x_u=1$).
As it is clear from eq. \eqref{vamanoeps}, 
again the expected value of $V_{us}$ manifests almost no dependence from the phase $\tilde{\beta}$ whereas it 
it is determined by the phase $\tilde{\delta}$. 
Conversely, $V_{cb}$ remains independent of the phase $\tilde{\delta}$.
However, 
as effect of the presence of the asymmetry, the rotation angle $s_{23}^d$ decreases while $s_{13}^d$ increases.
This modification causes the prediction of $|V_{cb}|$ to shift towards lower values.
As a result, an interval of values of the phase $\tilde{\beta}$ intercepts the experimental determination.
Furthermore, the asymmetry originates
a dependence of $|V_{ub}|$ on $\tilde{\delta}$ and of the ratio $|V_{ub}/V_{cb}|$ on $\tilde{\beta}$.
Similarly, the predictions of $V_{td}$ and $V_{ts}$ adjust to lower values.

Remarkably,
the values of $\tilde{\delta}$ and $\tilde{\beta}$ selected by $|V_{us}|$ and $|V_{cb}|$ %respectively
provide a prediction of all other observables which results within $1\sigma$ of their experimental constraints 
for any scale of new physics.
In fact, 
a $\chi^2$ fit  using the determinations of $|V_{us}|$, $|V_{ub}|$, $|V_{cb}|$ and $J$ 
(again we have $8$ parameters against $6$ eigenvalues plus $4$ CKM observables) 
returns in the minimum $\chi^2_\text{min}\approx 0$.
In figure \ref{asymm} we illustrate  
the new predictions (red bands) 
for Fritzsch-like texture at a scale between $10^3$~GeV (blue lines) and $10^{16}$~GeV (red lines),
confronted with the experimental determinations (grey bands) at $1\sigma$ confidence level.
We also indicate the $1\sigma$ interval of the phases $\tilde{\delta}$ and $\tilde{\beta}$ (green and yellow bands respectively).
In displaying the plots, the other variables to move inside the $1\sigma$ region.
The $1\sigma$ and $2\sigma$ confidence intervals of $\tilde{\delta}$ and $\tilde{\beta}$ 
are displayed in figure \ref{fig-phase}. 
In figure \ref{fig-vtdvts} we report the expectations on $V_{td}$ and $V_{ts}$ produced by these parameters.
It is evident that all these observables can be obtained within $1\sigma$ of the experimental constraints
independently on the scale of new physics.
We will now describe a more detailed analysis, 
allowing different values of the asymmetries 
in order to find the parameter space for which all observables are in agreement with the experimental constraints.

\subsubsection{Global numerical analysis}

In the general scenario in which both up and down Yukawa matrices can present the asymmetry, 
$x_d,x_u\neq 0$, we have $10$ conditions for $10$ parameters.
Therefore, 
we can wonder if an exact solution exists. 
We proceed as follows. We first evaluate the matrix entries $A_{d,u}$, $B_{d,u}$, $C_{d,u}$ in eq. \eqref{YFr} in terms of
the Yukawa ratios and of the parameters $x_d$, $x_u$, $\tilde{\beta}$, $\tilde{\delta}$.
Hence, we fit the CKM mixing elements. In particular,
we want to find the values of the asymmetry parameters $x_d,x_u$ and the phases $\tilde{\beta},\tilde{\delta}$
for which, if they exist, 
the mixing matrix in eq. \eqref{CKMF} can reproduce the $4$ independent quantities
describing the unitary matrix $V_\text{CKM}$ as determined by present global data reported in eqs. \eqref{angfit}, \eqref{Jgfit}.
Next, we want to investigate the $SU(5)$ motivated scenario with symmetric texture for up-type quarks, 
by imposing $x_u=1$. Hence, we perform
a $\chi^2$ fit of the four CKM observables with three parameters,
the asymmetry parameter $x_d$ and the phases $\tilde{\beta},\tilde{\delta}$.

In the following, we will use the central values of Yukawas ratios.
The ratios 
$m_u/m_{d}$ and $m_s/m_{d}$ are not well-known, as shown in figure \ref{mumd}.
For this reason, in principle 
$r_{ud}=m_u/m_d$ could be left as a parameter, so that we would obtain the functions
$\frac{y_d}{y_s} (r_{ud}) $, $\frac{y_u}{y_c} (r_{ud}) $,
$V_\text{CKM}(r_{ud},x_d,x_u, \tilde{\beta} , \tilde{\delta})$ which can vary with this ratio (as we did in the previous example).
However, since 
the central value of the determination 
$m_u/m_d$ 
turns out to be also a good point in our fit, we also 
impose $m_u/m_d=0.477$, in the central value.
We illustrate the analysis assuming that 
the Yukawa matrices acquire the Fritzsch form in eq. (\ref{YFr}) at the benchmark scales of
$10^3$~GeV, $10^6$~GeV, $10^{16}$~GeV.

\paragraph{\textbf{$\bm{\mu=10^3}$~GeV:}}
At $1$~TeV, we have (recalling that the ratios $ y_d/y_s$, $ y_u/y_c$ in eq. \eqref{input} remain
renormalization invariant)
\begin{align}
&  \frac{y_s}{y_b}=\frac{1}{53.21 }  \, , \qquad
\frac{y_c}{y_t}=\frac{1}{276.0}  \, , \qquad
 \frac{y_d}{y_b}=\frac{1}{1073}  \, , \qquad \frac{y_u}{y_t} =\frac{1}{1.37 \times 10^5} \, ,
 \label{input2}
\end{align}
corresponding to $y_c/y_t(\mu=1\, \text{TeV})\approx   (1-0.014) \, m_c/m_t$,
$ y_s/y_b(\mu=1\, \text{TeV})\approx     1.014  \, m_s/m_b$,
For the mixing with the third generation we find $V_{cb}(1\, \text{TeV})=1.014 \, V_{cb}$, 
$V_{ub}(1\, \text{TeV})=1.014 \, V_{ub}$, and the same for $V_{td}$ and $V_{ts}$.
After imposing theYukawas ratios \eqref{input2}, we can write the Yukawa matrices in terms of the four parameters 
$x_d,x_u,\tilde{\beta},\tilde{\delta}$.
Hence, we get
a system of four equations (we have to match three angles and $J$) to be solved with the four parameters
$x_d,x_u,\tilde{\beta},\tilde{\delta}$.
This system turns out to 
have a solution:
\begin{align}
& x_d = 3.15 \, , \quad  x_u =0.97  \, , \quad \tilde{\beta} = -0.75 \, , \quad \tilde{\delta} = -1.91 \, .
\label{sol1}
\end{align}
This means that the Fritzsch-like pattern in eq. \eqref{YFr} can be considered as a good flavour structure which
gives the right predictions of masses and mixings of quarks.
A second solution can be found with $(x_d,x_u,\tilde{\beta},\tilde{\delta})=(5.28,1.35,-1.25,3.09)$, 
which requires larger asymmetries and we are not going to consider here.

Given the result in eq. \eqref{sol1}, 
we are interested to analyse the scenario 
in which the up-quark Yukawa matrix assumes the original symmetric Fritzsch texture,
that is $x_u=1$. 
Having one less parameter,
we perform a $\chi^2$ fit of the three CKM angles and the CP-violating quantity $J$
with the parameters $x_d$, $\tilde{\delta}$, $\tilde{\beta}$.
We obtain $\chi^2_\text{min}=0.25$ in the minimum, 
with best fit values in: 
\begin{align}
& x_d = 3.16 \pm 0.17 \, , \quad  \tilde{\beta} = -0.78 \pm 0.02 \, , \quad \tilde{\delta} = -1.92\pm 0.04
\end{align}
where we also indicated the $1\sigma$ interval of the parameters ($\chi^2_\text{min}+1$).
We conclude that the canonical symmetric Fritzsch texture for up quarks is a good predictive ansatz
in models in which an asymmetry is generated in the mixing between the second and third generation in the down sector.

\paragraph{\textbf{$\bm{\mu=10^6}$~GeV:}}
Assuming $10^3$~TeV as the scale of new physics,
we have (with the ratios $ y_d/y_s$, $ y_u/y_c$ in eq. \eqref{input})
\begin{align}
&  \frac{y_s}{y_b}=\frac{1}{51.31 }  \, , \qquad
\frac{y_c}{y_t}=\frac{1}{286.2}  \, , \qquad
\frac{y_d}{y_b}  =\frac{1}{1034}  \, , \qquad \frac{y_u}{y_t} =\frac{1}{1.42 \times 10^5} \, .
 \label{input3}
\end{align}
corresponding to $y_c/y_t(10^3\, \text{TeV})\approx   0.95 \, m_c/m_t$ and
$ y_s/y_b(10^3\, \text{TeV})\approx     1.05  \, m_s/m_b$. 
For the mixing elements we find $V_{cb}(10^3\, \text{TeV})=1.051 \, V_{cb}$, 
$V_{ub}(10^3\, \text{TeV})=1.051 \, V_{ub}$ and the same for $V_{td}$ and $V_{ts}$.
By imposing the values of the Yukawas and the four CKM parameters we find the solution:
\begin{align}
& x_d = 3.09 \, , \quad  x_u =0.92  \, , \quad \tilde{\beta} = -0.75 \, , \quad \tilde{\delta} = -1.91 \, .
\end{align}
and a second solution in $(x_d,x_u,\tilde{\beta},\tilde{\delta})=(5.18,1.28,-1.25,3.09)$. 

Again we can fix $x_u=1$ so that the up-quark
Yukawa matrix assumes the original symmetric Fritzsch texture.
After imposing the Yukawa eigenvalues, we perform the fit of
the three CKM angles and $J$ with the parameters $x_d$, $\tilde{\delta}$, $\tilde{\beta}$.
We obtain  
$\chi^2_\text{min}=0.03$ in the minimum and the
$1\sigma$ intervals of parameters ($\chi^2_\text{min}+1$):
\begin{align}
& x_d = 3.14\pm 0.16 \, , \quad  \tilde{\beta} = -0.84\pm 0.02 \, , \quad \tilde{\delta} = -1.92\pm 0.04
\end{align}

\paragraph{\textbf{$\bm{\mu=10^{16}}$~GeV:}}
At $10^{16}$~GeV, we have (with the ratios $ y_d/y_s$, $ y_u/y_c$ in eq. \eqref{input})
\begin{align}
&  \frac{y_s}{y_b}=\frac{1}{48.40 }  \, , \qquad
\frac{y_c}{y_t}=\frac{1}{303.4}  \, , \qquad
\frac{y_d}{y_b} =\frac{1}{976}  \, , \qquad
\frac{y_u}{y_t} =\frac{1}{1.51 \times 10^5} \, .
 \label{input16}
\end{align}
meaning $y_c/y_t(10^{16}\, \text{GeV})\approx   (1-0.1) \, m_c/m_t$,
$ y_s/y_b(10^{16}\, \text{GeV})\approx     1.1  \, m_s/m_b$.
For the mixing elements we have $V_{cb}(10^{16}\, \text{GeV})=1.114 \, V_{cb}$, 
$V_{ub}(10^{16}\, \text{GeV})=1.114 \, V_{ub}$. %
We find the solution
\begin{align}
& x_d = 3.00 \, , \quad  x_u =0.85  \, , \quad \tilde{\beta} = -0.75 \, , \quad \tilde{\delta} = -1.91 \, .
\end{align}
and a second solution in $(x_d,x_u,\tilde{\beta},\tilde{\delta})=(5.03,1.17,-1.25,3.09)$, with larger asymmetries.

By imposing the condition $x_u=1$, we perform the $\chi^2$ fit
of the three CKM angle and invariant $J$. We receive the minimum $\chi^2_\text{min}=0.75$ in the best fit 
values of the parameters (and relative $1\sigma$ interval, $\chi^2_\text{min}+1$) 
\begin{align}
& x_d = 3.12\pm 0.16 \, , \quad  \tilde{\beta} = -0.93 \pm 0.02 \, , \quad \tilde{\delta} = -1.94 \pm 0.04
\end{align}

\medskip
\paragraph{Summary}

\begin{figure}
\centering
\includegraphics[width=0.32\textwidth]{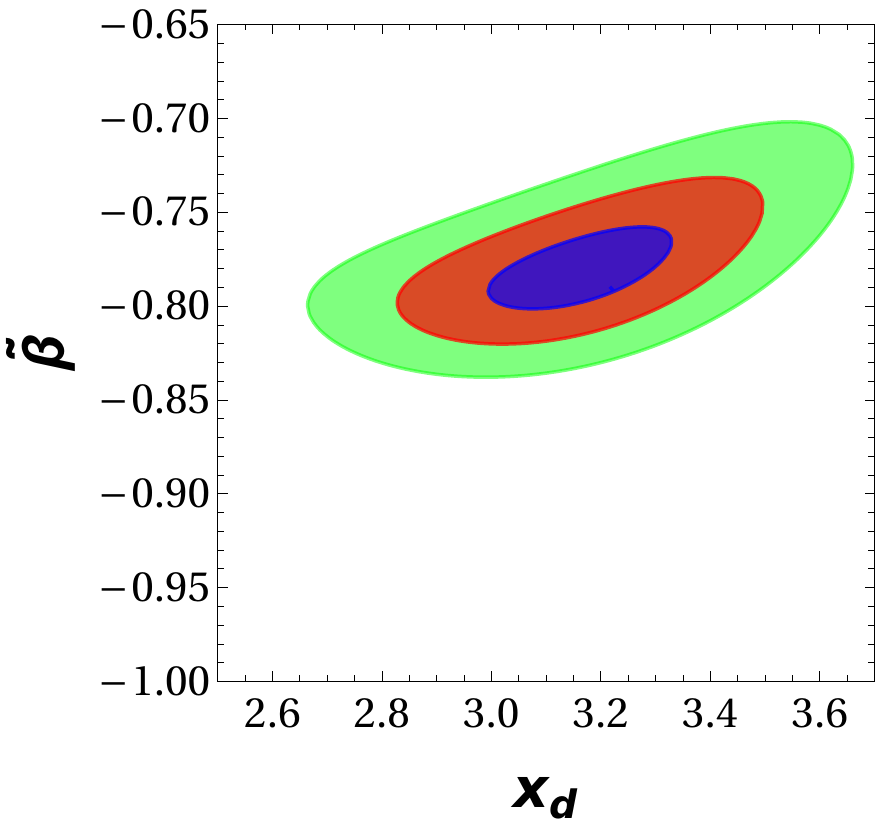}
\includegraphics[width=0.32\textwidth]{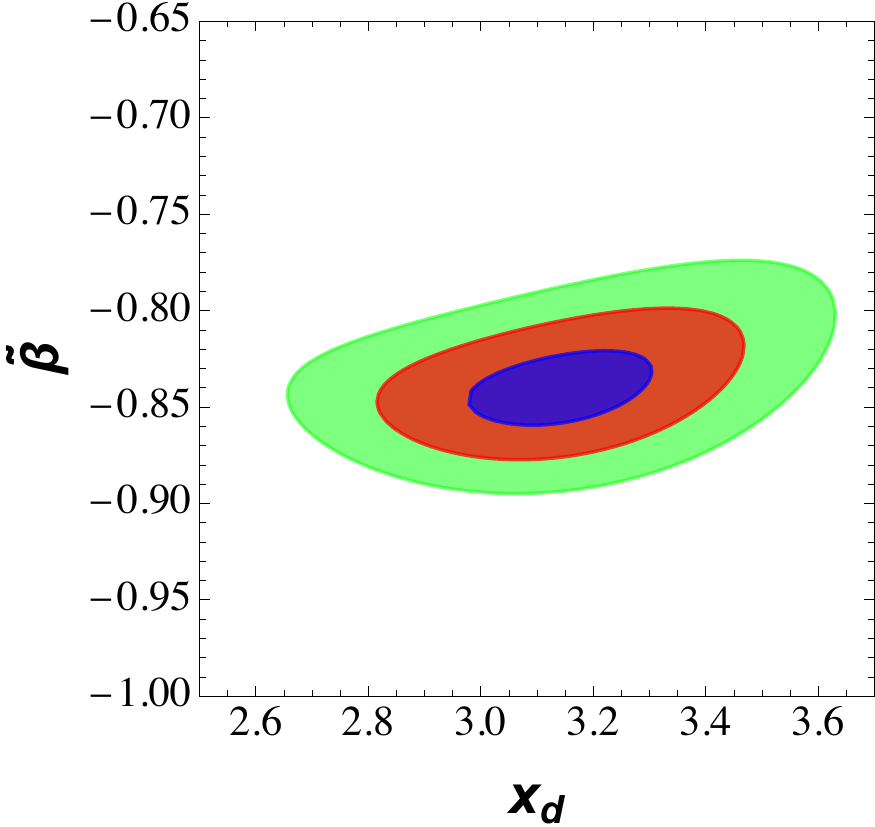}
\includegraphics[width=0.32\textwidth]{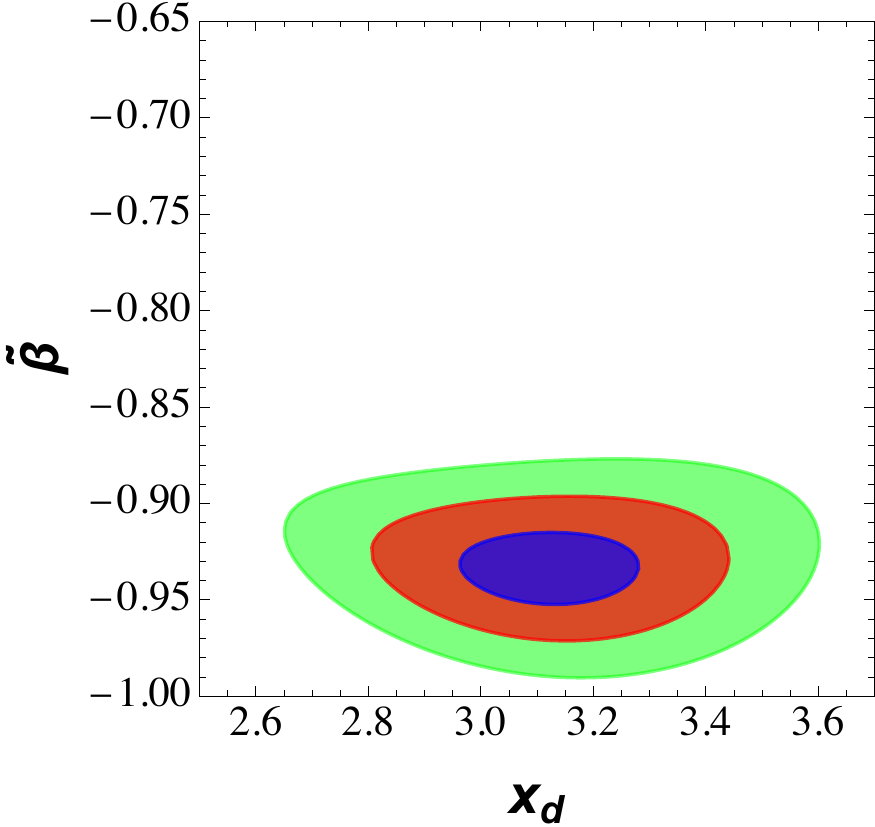}
 \\ \vspace{10pt}
\includegraphics[width=0.32\textwidth]{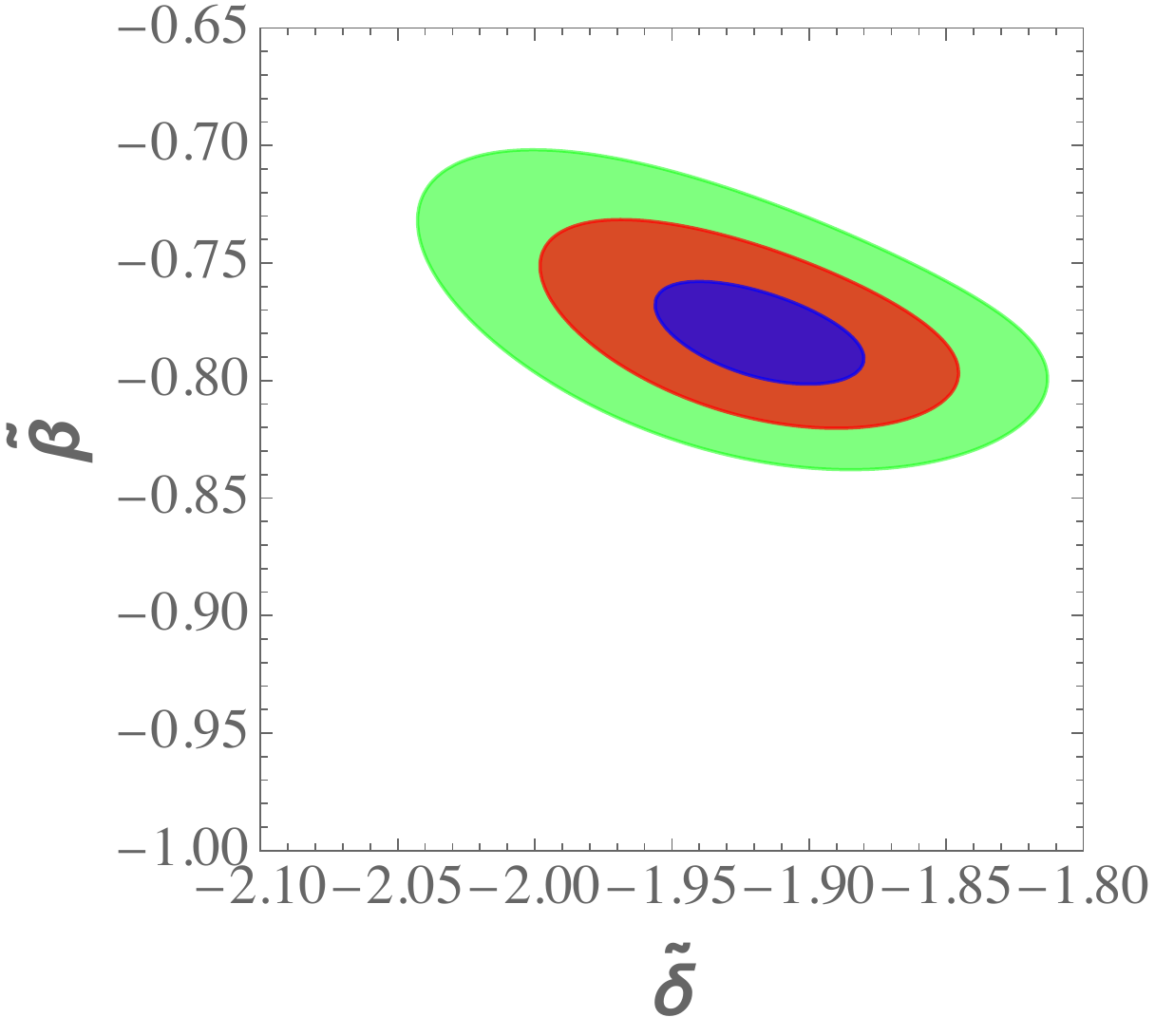}
\includegraphics[width=0.32\textwidth]{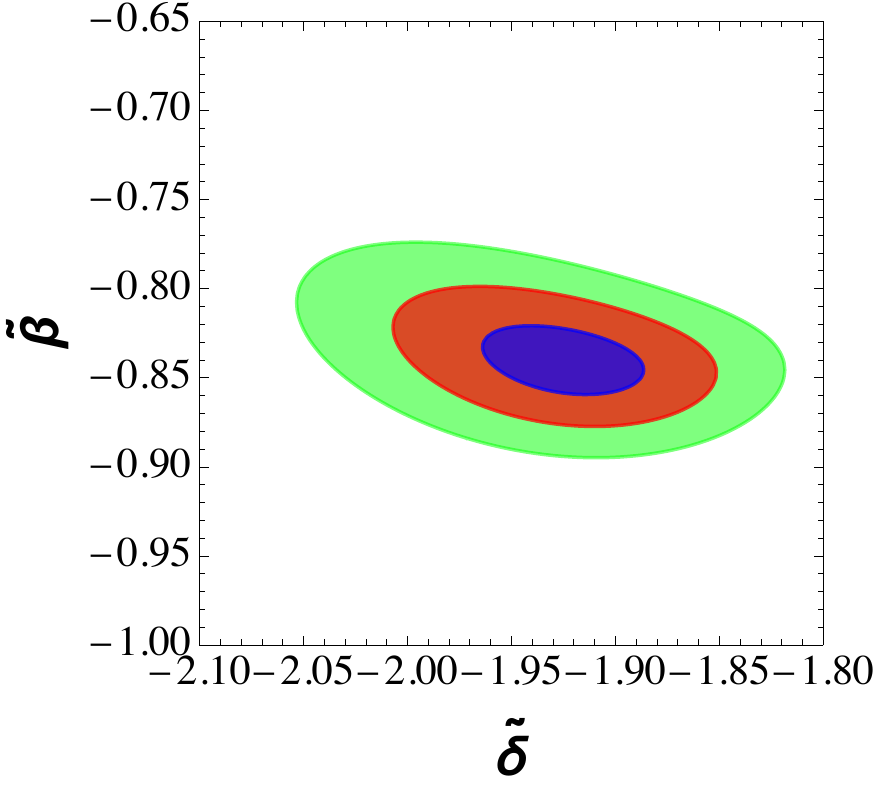}
\includegraphics[width=0.32\textwidth]{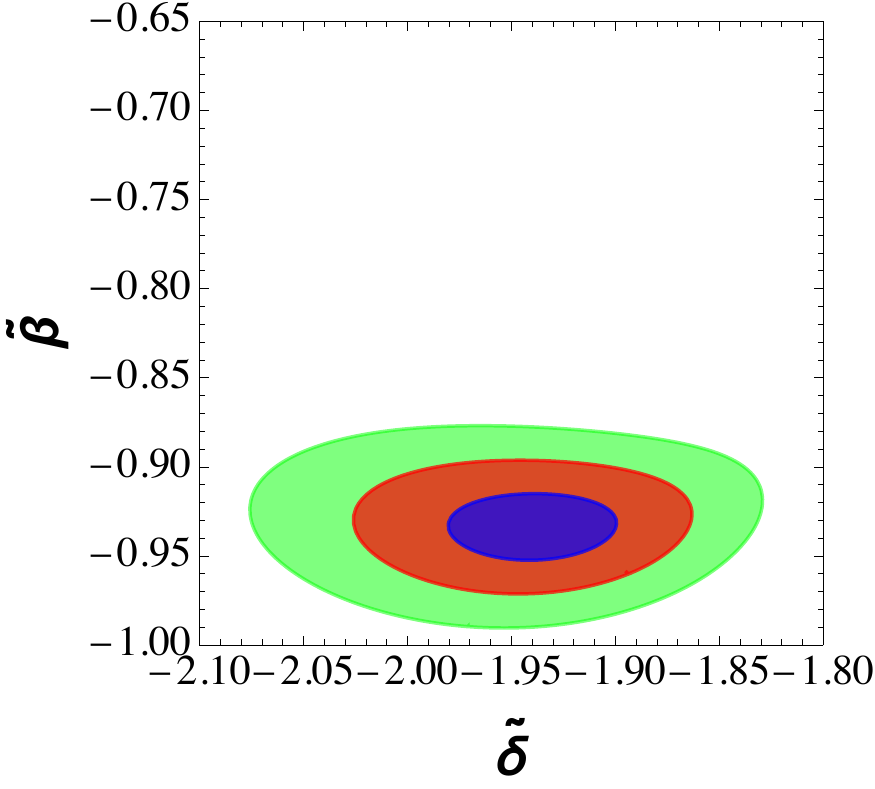}
 \\ \vspace{10pt}
\includegraphics[width=0.32\textwidth]{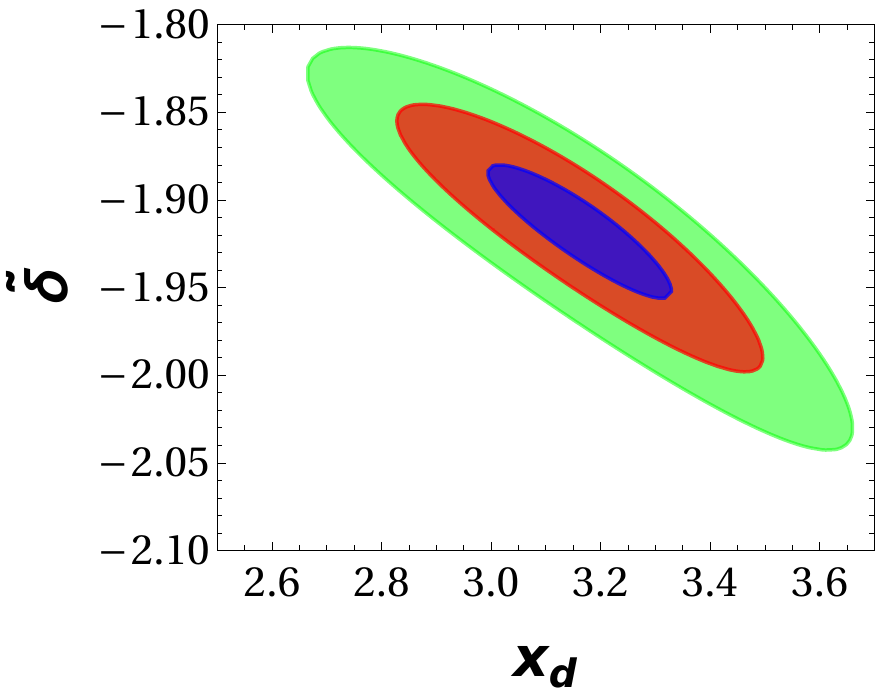}
\includegraphics[width=0.32\textwidth]{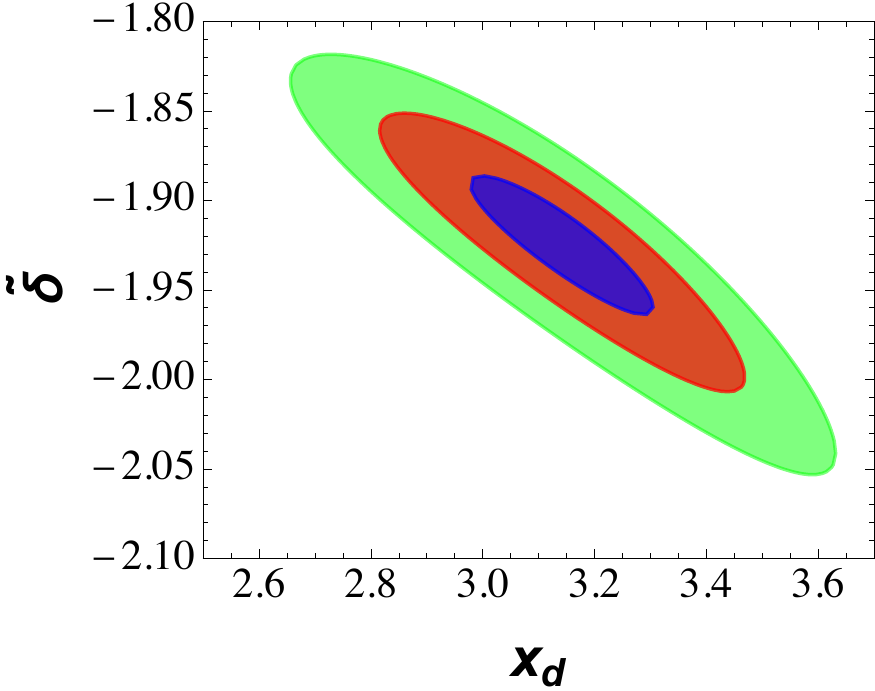}
\includegraphics[width=0.32\textwidth]{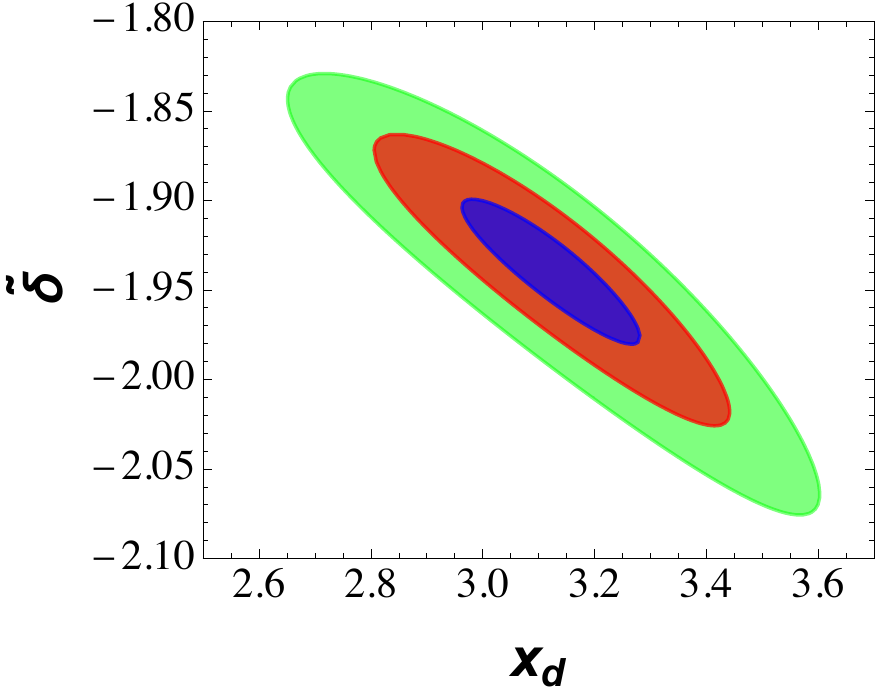}
\caption{\label{global-proj} 
Parameter space in the scenario with $x_u=1$ (symmetric Fritzsch texture for up-type quarks, 
asymmetric 23 entries 
for down-type quarks, see eq. \eqref{YFr})
in the $x_d$-$\tilde{\beta}$, $\tilde{\delta}$-$\tilde{\beta}$ and $x_d$-$\tilde{\beta}$ planes,
marginalizing over the other variable.
$1\sigma$, $2\sigma$ and $3\sigma$ preferred regions of 
the parameters are indicated ($\chi^2_\text{min}+1$, $\chi^2_\text{min}+4$, $\chi^2_\text{min}+9$),
assuming Yukawa matrices of Fritzsch-like form at $10^3$~GeV (left), $10^6$~TeV (centre), $10^{16}$~TeV (right). 
}
\end{figure}

We conclude that a symmetric Fritzsch texture for up quarks ($x_u\approx 1$) and a minimally modified Fritzsch texture
as in eq. \eqref{YFr} for down quarks with $x_d\approx 3$ are good flavour structures,
which can predict the right masses of quarks as well as CKM mixings and phase.
In particular, $3\lesssim x_d \lesssim 3.3$ is a good interval for any energy scale at which the Fritzsch-like textures exist.
In figure \ref{global-proj} we present the results of the analysis in the $x_d$-$\tilde{\beta}$, 
$\tilde{\delta}$-$\tilde{\beta}$ and $x_d$-$\tilde{\beta}$
planes in the scenario with $x_u=1$,
marginalizing over the other variable, assuming Fritzsch-like texture at $10^3$~GeV, $10^6$~GeV, $10^{16}$~GeV
(left, centre and right respectively).
We show the $1\sigma$, $2\sigma$, $3\sigma$ (blue, red and green regions) confidence intervals %of the total fit 
($\chi^2_\text{min}+1$, $\chi^2_\text{min}+4$, $\chi^2_\text{min}+9$). 
The best fit values give for the three benchmark scales respectively the magnitudes of CKM elements:
\begin{align}
& 
\left(\!
\begin{array}{ccc}
0.97435 & 0.22500 & 0.00369 \\
0.22486 & 0.97347  & 0.0418 \\
0.00865 & 0.0410 & 0.99910
\end{array}
\! \right) , \,  
\left(\!
\begin{array}{ccc}
0.97435 & 0.22500 & 0.00367 \\
0.22485 & 0.97341  & 0.0416 \\
0.00880 & 0.0409 & 0.99903
\end{array}
\! \right)
, \,  
\left(\!
\begin{array}{ccc}
0.97435 & 0.22500 & 0.00365 \\
0.22481 & 0.97330  & 0.0415 \\
0.00901 & 0.0407 & 0.99892
\end{array}
\! \right)
\end{align}
in perfect agreement with the observables in eq. \eqref{globalckm}.

By adopting different choices of experimental results, e.g. %By employing 
some of the determinations in table \ref{values} instead of the global fit, 
the result of the numerical analysis would be very similar.
In fact, we find that 
this flavour pattern is able to reproduce all CKM observables within $1 \sigma$.

%%%%%%%%%%%%%%%%%%%%%%%%%%%%%%%%%%%%%%%%%%%%%%%%%%%%%%%%%%%%%%

%%%%%%%%%%%%%%%%%%%%%%%%%%%%%%%%%%%%%%%%%%%%%%%%%%

\section{Conclusion}
\label{conclusion}

The hierarchy between fermion masses, their mixing pattern as well as the replication of families itself 
remain a mystery in the context of the Standard Model or grand unified theories.
It is intriguing to think that clues for an explanation may be found in the existing
relations between mass ratios and mixing angles, 
as the formula for the Cabibbo angle $V_{us} = \sqrt{m_d/m_s}$,
which may be regarded as 
not accidental 
but rather connected to an underlying flavour theory.
These relations can be predicted by  
mass matrices with reduced number of parameters.
Moreover, present experimental data and lattice computations have reached enough precision to scrutinize
some of the hypothesis on the Yukawa textures. 

In this work,
we concentrated on the predictive Fritzsch texture for quark masses, with $6$ zero entries 
(three in the up-quarks Yukawa matrix and three in the down-type one).
This flavour structure contains $8$ parameters which should match $10$ observables: six quarks masses,
three mixing angles and one CP-violating phase.
However, the original symmetric ansatz is excluded by present data. 
Nevertheless, an asymmetry in the mixing between the second and third generation 
can be introduced.
We analyzed 
this asymmetric version of the Fritzsch texture, with the same vanishing elements,
considering its possible origin and confronting its predictions with recent precise experimental and lattice results
on quark masses and mixings.

In particular, we showed that the canonical symmetric Fritzsch form for up-type quarks 
can be combined with the 
asymmetric texture for down quarks. 
In this way, with $9$ parameters to match $10$ observables,
all values of mass ratios and 
CKM matrix observables
can be reproduced within $\sim 1 \sigma$,
independently on the energy scale at which the Fritzsch structure is generated. 

We showed how this flavour pattern,
the symmetric texture for up quarks and the asymmetric one for down quarks,
can naturally arise from models with $SU(3)_H$ gauge family symmetry
in the context of Standard Model or grand unified theories.

\appendix

\acknowledgments

We would like to thank Paolo Panci, Robert Ziegler and 
Antonio Rodríguez Sánchez for useful discussions.
The work of Z.B. was supported in part by Ministero dell’Istruzione, Universit\'a e della Ricerca (MIUR)
under the program PRIN 2017, Grant 2017X7X85K ``The dark universe: synergic multimessenger approach”.
We would like to remember Harald Fritzsch who recently passed away
for his pioneering contribution towards the understanding of flavour structures.


\begin{thebibliography}{99}

%\cite{Cabibbo:1963yz}
\bibitem{Cabibbo:1963yz}
N.~Cabibbo,
``Unitary Symmetry and Leptonic Decays,''
Phys. Rev. Lett. \textbf{10}, 531-533 (1963)
doi:10.1103/PhysRevLett.10.531
%7275 citations counted in INSPIRE as of 19 Apr 2023

%\cite{Kobayashi:1973fv}
\bibitem{Kobayashi:1973fv}
M.~Kobayashi and T.~Maskawa,
``CP Violation in the Renormalizable Theory of Weak Interaction,''
Prog. Theor. Phys. \textbf{49}, 652-657 (1973)
doi:10.1143/PTP.49.652
%11488 citations counted in INSPIRE as of 19 Apr 2023

%%\cite{Chau:1984fp}
%\bibitem{Chau:1984fp}
%L.~L.~Chau and W.~Y.~Keung,
%``Comments on the Parametrization of the Kobayashi-Maskawa Matrix,''
%Phys. Rev. Lett. \textbf{53}, 1802 (1984)
%doi:10.1103/PhysRevLett.53.1802
%%1066 citations counted in INSPIRE as of 19 Apr 2023

%\cite{Workman:2022ynf}
\bibitem{PDG22}
R.~L.~Workman \textit{et al.} [Particle Data Group],
``Review of Particle Physics,''
PTEP \textbf{2022}, 083C01 (2022)
doi:10.1093/ptep/ptac097
%3 citations counted in INSPIRE as of 12 Jul 2022

%\cite{Jarlskog:1985ht}
\bibitem{Jarlskog:1985ht}
C.~Jarlskog,
``Commutator of the Quark Mass Matrices in the Standard Electroweak Model and a Measure of Maximal $CP$~Nonconservation,''
Phys. Rev. Lett. \textbf{55}, 1039 (1985)
doi:10.1103/PhysRevLett.55.1039
%1968 citations counted in INSPIRE as of 19 Apr 2023


%\cite{Weinberg:1977hb}
\bibitem{Weinberg1977}
S.~Weinberg,
``The Problem of Mass,''
Trans. New York Acad. Sci. \textbf{38}, 185-201 (1977)
doi:10.1111/j.2164-0947.1977.tb02958.x
%408 citations counted in INSPIRE as of 30 Nov 2021

%\cite{Wilczek:1977uh}
\bibitem{Wilczek1977}
F.~Wilczek and A.~Zee,
``Discrete Flavor Symmetries and a Formula for the Cabibbo Angle,''
Phys. Lett. B \textbf{70}, 418 (1977)
[erratum: Phys. Lett. B \textbf{72}, 504 (1978)]
doi:10.1016/0370-2693(77)90403-8
%299 citations counted in INSPIRE as of 30 Nov 2021

%\cite{Fritzsch:1977za}
\bibitem{Fritzsch1977}
H.~Fritzsch,
``Calculating the Cabibbo Angle,''
Phys. Lett. B \textbf{70}, 436-440 (1977)
doi:10.1016/0370-2693(77)90408-7
%593 citations counted in INSPIRE as of 30 Nov 2021

%\cite{Fritzsch:1977vd}
\bibitem{Fritzsch78}
H.~Fritzsch,
``Weak Interaction Mixing in the Six - Quark Theory,''
Phys. Lett. B \textbf{73}, 317-322 (1978)
doi:10.1016/0370-2693(78)90524-5
%868 citations counted in INSPIRE as of 31 Aug 2021

%\cite{Fritzsch:1979zq}
\bibitem{Fritzsch79}
H.~Fritzsch,
``Quark Masses and Flavor Mixing,''
Nucl. Phys. B \textbf{155}, 189-207 (1979)
doi:10.1016/0550-3213(79)90362-6
%725 citations counted in INSPIRE as of 31 Aug 2021

%\cite{Fritzsch:1983dc}
\bibitem{Fritzsch83}
H.~Fritzsch,
``Flavor mixing and the masses of leptons and quarks,"
%``FLAVOR MIXING AND THE MASSES OF LEPTONS AND QUARKS,''
J. Phys. Colloq. \textbf{45}, no.C3, 189-196 (1984)
doi:10.1051/jphyscol:1984332


%\cite{Glashow:1970gm}
\bibitem{GIM}
S.~L.~Glashow, J.~Iliopoulos and L.~Maiani,
``Weak Interactions with Lepton-Hadron Symmetry,''
Phys. Rev. D \textbf{2}, 1285-1292 (1970)
doi:10.1103/PhysRevD.2.1285

%\cite{Glashow:1976nt}
\bibitem{Weinberg}
S.~L.~Glashow and S.~Weinberg,
``Natural Conservation Laws for Neutral Currents,''
Phys. Rev. D \textbf{15}, 1958 (1977)
doi:10.1103/PhysRevD.15.1958

%\cite{Paschos:1976ay}
\bibitem{Paschos}
E.~A.~Paschos,
``Diagonal Neutral Currents,''
Phys. Rev. D \textbf{15}, 1966 (1977)
doi:10.1103/PhysRevD.15.1966


%\cite{Gatto:1978dy}
\bibitem{Gatto}
R.~Gatto, G.~Morchio and F.~Strocchi,
``Natural Flavor Conservation in the Neutral Currents and the Determination of the Cabibbo Angle,''
Phys. Lett. B \textbf{80}, 265-268 (1979)
doi:10.1016/0370-2693(79)90213-2

 %\cite{Berezhiani:1983hm}
\bibitem{PLB}
Z.~G.~Berezhiani,
``The Weak Mixing Angles in Gauge Models with Horizontal Symmetry: A New Approach to Quark and Lepton Masses,''
Phys. Lett. B \textbf{129}, 99-102 (1983)
doi:10.1016/0370-2693(83)90737-2


%\cite{Berezhiani:1985in}
\bibitem{PLB2}
Z.~G.~Berezhiani,
``Horizontal Symmetry and Quark - Lepton Mass Spectrum: The $SU(5) \times SU(3)_H$ Model,''
Phys. Lett. B \textbf{150}, 177-181 (1985)
doi:10.1016/0370-2693(85)90164-9

%\cite{Kang:1992eq}
\bibitem{Kang}
K.~Kang, J.~Flanz and E.~Paschos,
``Confronting experiments with numerical analysis of the Fritzsch type mass matrices,''
Z. Phys. C \textbf{55}, 75-82 (1992)
doi:10.1007/BF01558290
%17 citations counted in INSPIRE as of 31 Aug 2021

%\cite{Ramond:1993kv}
\bibitem{Ramond:1993kv}
P.~Ramond, R.~G.~Roberts and G.~G.~Ross,
``Stitching the Yukawa quilt,''
Nucl. Phys. B \textbf{406}, 19-42 (1993)
doi:10.1016/0550-3213(93)90159-M
[arXiv:hep-ph/9303320 [hep-ph]].
%423 citations counted in INSPIRE as of 19 Apr 2023

%\cite{Berezhiani:1991tj}
\bibitem{Lavoura}
Z.~G.~Berezhiani and L.~Lavoura,
``Fritzsch like model for the quark mass matrices with a large first - third generation mixing,''
Phys. Rev. D \textbf{45}, 934-945 (1992)
doi:10.1103/PhysRevD.45.934

%\cite{Giraldo:2011ya}
\bibitem{Giraldo:2011ya}
Y.~Giraldo,
``Texture Zeros and WB Transformations in the Quark Sector of the Standard Model,''
Phys. Rev. D \textbf{86}, 093021 (2012)
doi:10.1103/PhysRevD.86.093021
[arXiv:1110.5986 [hep-ph]].
%17 citations counted in INSPIRE as of 18 Apr 2023


%\cite{Xing:2015sva}
\bibitem{Xing:2015sva}
Z.~z.~Xing and Z.~h.~Zhao,
``On the four-zero texture of quark mass matrices and its stability,''
Nucl. Phys. B \textbf{897}, 302-325 (2015)
doi:10.1016/j.nuclphysb.2015.05.027
[arXiv:1501.06346 [hep-ph]].
%28 citations counted in INSPIRE as of 18 Apr 2023

%\cite{Linster:2018avp}
\bibitem{Linster:2018avp}
M.~Linster and R.~Ziegler,
``A Realistic $U(2)$ Model of Flavor,''
JHEP \textbf{08}, 058 (2018)
doi:10.1007/JHEP08(2018)058
[arXiv:1805.07341 [hep-ph]].
%35 citations counted in INSPIRE as of 18 Apr 2023

%\cite{Bagai:2021nsl}
\bibitem{Bagai}
A.~Bagai, A.~Vashisht, N.~Awasthi, G.~Ahuja and M.~Gupta,
``Probing texture 4 zero quark mass matrices in the era of precision measurements,''
[arXiv:2110.05065 [hep-ph]].
%2 citations counted in INSPIRE as of 30 Nov 2021

%\cite{Fritzsch:2021ipb}
\bibitem{Fritzsch2021}
H.~Fritzsch, Z.~z.~Xing and D.~Zhang,
``Correlations between quark mass and flavor mixing hierarchies,''
Nucl. Phys. B \textbf{974}, 115634 (2022)
doi:10.1016/j.nuclphysb.2021.115634
[arXiv:2111.06727 [hep-ph]].
%6 citations counted in INSPIRE as of 28 Dec 2022

%\cite{Berezhiani:1998vn}
\bibitem{Rossi1}
Z.~Berezhiani and A.~Rossi,
``Grand unified textures for neutrino and quark mixings,''
JHEP \textbf{03}, 002 (1999)
doi:10.1088/1126-6708/1999/03/002
[arXiv:hep-ph/9811447 [hep-ph]].

%\cite{Roberts:2001zy}
\bibitem{Roberts:2001zy}
R.~G.~Roberts, A.~Romanino, G.~G.~Ross and L.~Velasco-Sevilla,
``Precision Test of a Fermion Mass Texture,''
Nucl. Phys. B \textbf{615}, 358-384 (2001)
doi:10.1016/S0550-3213(01)00408-4
[arXiv:hep-ph/0104088 [hep-ph]].
%147 citations counted in INSPIRE as of 19 Apr 2023
 
  %\cite{Froggatt:1978nt}
\bibitem{Froggatt}
C.~D.~Froggatt and H.~B.~Nielsen,
``Hierarchy of Quark Masses, Cabibbo Angles and CP Violation,''
Nucl. Phys. B \textbf{147}, 277-298 (1979).
doi:10.1016/0550-3213(79)90316-X

%\cite{Ibanez:1994ig}
\bibitem{Ibanez}
L.~E.~Ibanez and G.~G.~Ross,
``Fermion masses and mixing angles from gauge symmetries,''
Phys. Lett. B \textbf{332}, 100-110 (1994)
doi:10.1016/0370-2693(94)90865-6
[arXiv:hep-ph/9403338 [hep-ph]].

%\cite{Binetruy:1996xk}
\bibitem{Binetruy}
P.~Binetruy, S.~Lavignac and P.~Ramond,
``Yukawa textures with an anomalous horizontal Abelian symmetry,''
Nucl. Phys. B \textbf{477}, 353-377 (1996)
doi:10.1016/0550-3213(96)00296-9
[arXiv:hep-ph/9601243 [hep-ph]].

%\cite{Dudas:1996fe}
\bibitem{Dudas}
E.~Dudas, C.~Grojean, S.~Pokorski and C.~A.~Savoy,
``Abelian flavor symmetries in supersymmetric models,''
Nucl. Phys. B \textbf{481}, 85-108 (1996)
doi:10.1016/S0550-3213(96)90123-6
[arXiv:hep-ph/9606383 [hep-ph]].

%\cite{Berezhiani:1996nu}
\bibitem{Tavartkiladze1}
Z.~Berezhiani and Z.~Tavartkiladze,
``Anomalous U(1) symmetry and missing doublet SU(5) model,''
Phys. Lett. B \textbf{396}, 150-160 (1997)
doi:10.1016/S0370-2693(97)00122-6
[arXiv:hep-ph/9611277 [hep-ph]].

%\cite{Berezhiani:1996bv}
\bibitem{Tavartkiladze2}
Z.~Berezhiani and Z.~Tavartkiladze,
``More missing VEV mechanism in supersymmetric SO(10) model,''
Phys. Lett. B \textbf{409}, 220-228 (1997)
doi:10.1016/S0370-2693(97)00873-3
[arXiv:hep-ph/9612232 [hep-ph]].

%\cite{Grimus:2004hf}
\bibitem{Grimus:2004hf}
W.~Grimus, A.~S.~Joshipura, L.~Lavoura and M.~Tanimoto,
``Symmetry realization of texture zeros,''
Eur. Phys. J. C \textbf{36}, 227-232 (2004)
doi:10.1140/epjc/s2004-01896-y
[arXiv:hep-ph/0405016 [hep-ph]].
%171 citations counted in INSPIRE as of 19 Apr 2023

%\cite{Ferreira:2010ir}
\bibitem{Ferreira:2010ir}
P.~M.~Ferreira and J.~P.~Silva,
``Abelian symmetries in the two-Higgs-doublet model with fermions,''
Phys. Rev. D \textbf{83}, 065026 (2011)
doi:10.1103/PhysRevD.83.065026
[arXiv:1012.2874 [hep-ph]].
%39 citations counted in INSPIRE as of 18 Apr 2023
%\cite{Serodio:2013gka}
\bibitem{Serodio:2013gka}
H.~Ser\^odio,
``Yukawa sector of Multi Higgs Doublet Models in the presence of Abelian symmetries,''
Phys. Rev. D \textbf{88}, no.5, 056015 (2013)
doi:10.1103/PhysRevD.88.056015
[arXiv:1307.4773 [hep-ph]].
%27 citations counted in INSPIRE as of 18 Apr 2023
%\cite{Bjorkeroth:2018ipq}
\bibitem{Bjorkeroth:2018ipq}
F.~Bj\"orkeroth, L.~Di Luzio, F.~Mescia and E.~Nardi,
``$U(1)$ flavour symmetries as Peccei-Quinn symmetries,''
JHEP \textbf{02}, 133 (2019)
doi:10.1007/JHEP02(2019)133
[arXiv:1811.09637 [hep-ph]].
%24 citations counted in INSPIRE as of 19 Apr 2023
 

%\cite{Berezhiani:1982rr}
\bibitem{Chkareuli} 
Z.~G.~Berezhiani and J.~L.~Chkareuli,
``Mass Of The T Quark And The Number Of Quark Lepton Generations'',
JETP Lett. \textbf{35}, 612-615 (1982)
%[erratum: JETP Lett. \textbf{36}, 380 (1982); erratum: Pisma Zh. Eksp. Teor. Fiz. \textbf{36}, 312 (1982)]
%67 citations counted in INSPIRE as of 31 Aug 2021

%\cite{Berezhiani:1983rk}
\bibitem{Chkareuli2}
Z.~G.~Berezhiani and J.~L.~Chkareuli,
``Quark-leptonic families in a model with $SU(5) \times SU(3)$ symmetry"
%``QUARK - LEPTONIC FAMILIES IN A MODEL WITH SU(5) X SU(3) SYMMETRY. (IN RUSSIAN),''
Sov. J. Nucl. Phys. \textbf{37}, 618-626 (1983)

%\cite{Berezhiani:1985vxy}
\bibitem{Chkareuli3}
Z.~G.~Berezhiani and J.~L.~Chkareuli,
``Horizontal Symmetry: Masses and Mixing Angles of Quarks and Leptons of Different Generations: 
Neutrino Mass and Neutrino Oscillation,''
Sov. Phys. Usp. \textbf{28}, 104-105 (1985)
doi:10.1070/PU1985v028n01ABEH003846

%\cite{Berezhiani:1996kk}
\bibitem{Berezhiani:1996kk}
Z.~Berezhiani,
``Problem of flavor in SUSY GUT and horizontal symmetry,''
Nucl. Phys. B Proc. Suppl. \textbf{52}, 153-158 (1997)
doi:10.1016/S0920-5632(96)00552-X
[arXiv:hep-ph/9607363 [hep-ph]].

%\cite{Berezhiani:2000cg}
\bibitem{Berezhiani:2000cg}
Z.~Berezhiani and A.~Rossi,
``Predictive grand unified textures for quark and neutrino masses and mixings,''
Nucl. Phys. B \textbf{594}, 113-168 (2001)
doi:10.1016/S0550-3213(00)00653-2
[arXiv:hep-ph/0003084 [hep-ph]].
%102 citations counted in INSPIRE as of 05 Dec 2021

%\cite{Belfatto:2018cfo}
\bibitem{Belfatto:2018cfo}
B.~Belfatto and Z.~Berezhiani,
``How light the lepton flavor changing gauge bosons can be,''
Eur. Phys. J. C \textbf{79}, no.3, 202 (2019)
doi:10.1140/epjc/s10052-019-6724-5
[arXiv:1812.05414 [hep-ph]]

 %\cite{Georgi:1974sy}
\bibitem{Georgi}
H.~Georgi and S.~L.~Glashow,
``Unity of All Elementary Particle Forces,''
Phys. Rev. Lett. \textbf{32}, 438-441 (1974)
doi:10.1103/PhysRevLett.32.438

%\cite{Berezhiani:1996ii}
\bibitem{Berezhiani:1996ii}
Z.~Berezhiani,
``Unified picture of the particle and sparticle masses in SUSY GUT,''
Phys. Lett. B \textbf{417}, 287-296 (1998)
doi:10.1016/S0370-2693(97)01359-2
[arXiv:hep-ph/9609342 [hep-ph]]
%101 citations counted in INSPIRE as of 30 Nov 2021

%\cite{Berezhiani:2003xm}
\bibitem{IJMPA}
Z.~Berezhiani,
``Mirror world and its cosmological consequences,''
Int. J. Mod. Phys. A \textbf{19}, 3775-3806 (2004)
doi:10.1142/S0217751X04020075
[arXiv:hep-ph/0312335 [hep-ph]] 

%\cite{Berezhiani:2005ek}
\bibitem{Alice}
Z.~Berezhiani,
``Through the looking-glass: Alice's adventures in mirror world,''
doi:10.1142/9789812775344\_0055
[arXiv:hep-ph/0508233 [hep-ph]]

%\cite{Berezhiani:1995tr}
\bibitem{ICTP}
Z.~Berezhiani,
``Fermion masses and mixing in SUSY GUT,''
ICTP Summer School in High-energy Physics and Cosmology, 
[arXiv:hep-ph/9602325 [hep-ph]].

%\cite{Berezhiani:1990wn}
\bibitem{Khlopov1}
Z.~G.~Berezhiani and M.~Y.~Khlopov,
``The Theory of broken gauge symmetry of families,''
Sov. J. Nucl. Phys. \textbf{51}, 739-746 (1990)
%121 citations counted in INSPIRE as of 05 Dec 2021 


%\cite{Berezhiani:1990jj}
\bibitem{Khlopov2}
Z.~G.~Berezhiani and M.~Y.~Khlopov,
``Physical and astrophysical consequences of breaking of the symmetry of families,''
Sov. J. Nucl. Phys. \textbf{51}, 935-942 (1990)
%76 citations counted in INSPIRE as of 05 Dec 2021

%\cite{Berezhiani:1990sy}
\bibitem{Khlopov3}
Z.~G.~Berezhiani and M.~Y.~Khlopov,
``Physics of cosmological dark matter in the theory of broken family symmetry,''
Sov. J. Nucl. Phys. \textbf{52}, 60-64 (1990)
%56 citations counted in INSPIRE as of 05 Dec 2021

%\cite{Berezhiani:1989fp}
\bibitem{Khlopov4}
Z.~G.~Berezhiani and M.~Y.~Khlopov,
``Cosmology of Spontaneously Broken Gauge Family Symmetry,''
Z. Phys. C \textbf{49}, 73-78 (1991)
doi:10.1007/BF01570798
%159 citations counted in INSPIRE as of 05 Dec 2021

%\cite{Berezhiani:1989fs}
\bibitem{Homeriki1}
Z.~G.~Berezhiani, M.~Y.~Khlopov and R.~R.~Khomeriki,
``On the Possible Test of Quantum Flavor Dynamics in the Searches for Rare Decays of Heavy Particles,''
Sov. J. Nucl. Phys. \textbf{52}, 344-347 (1990)
%FERMILAB-PUB-89-204-A.
%32 citations counted in INSPIRE as of 05 Dec 2021

%\cite{Berezhiani:1989fu}
\bibitem{Homeriki2}
Z.~G.~Berezhiani, M.~Y.~Khlopov and R.~R.~Khomeriki,
``Cosmic Nonthermal Electromagnetic Background from Axion Decays in the Models with Low Scale of Family Symmetry Breaking,''
Sov. J. Nucl. Phys. \textbf{52}, 65-68 (1990)
%FERMILAB-PUB-89-203-A.

%\cite{Berezhiani:1992rk}
\bibitem{Sakharov}
Z.~G.~Berezhiani, A.~S.~Sakharov and M.~Y.~Khlopov,
``Primordial background of cosmological axions,''
Sov. J. Nucl. Phys. \textbf{55}, 1063-1071 (1992)

%\cite{DiLuzio:2017ogq}
\bibitem{DiLuzio:2017ogq}
L.~Di Luzio, F.~Mescia, E.~Nardi, P.~Panci and R.~Ziegler,
``Astrophobic Axions,''
Phys. Rev. Lett. \textbf{120}, no.26, 261803 (2018)
doi:10.1103/PhysRevLett.120.261803
[arXiv:1712.04940 [hep-ph]].
%59 citations counted in INSPIRE as of 18 Apr 2023

%\cite{DiLuzio:2019mie}
\bibitem{DiLuzio:2019mie}
L.~Di Luzio,
``Flavour Violating Axions,''
EPJ Web Conf. \textbf{234}, 01005 (2020)
doi:10.1051/epjconf/202023401005
[arXiv:1911.02591 [hep-ph]].
%2 citations counted in INSPIRE as of 18 Apr 2023

%\cite{MartinCamalich:2020dfe}
\bibitem{MartinCamalich:2020dfe}
J.~Martin Camalich, M.~Pospelov, P.~N.~H.~Vuong, R.~Ziegler and J.~Zupan,
``Quark Flavor Phenomenology of the QCD Axion,''
Phys. Rev. D \textbf{102}, no.1, 015023 (2020)
doi:10.1103/PhysRevD.102.015023
[arXiv:2002.04623 [hep-ph]].
%87 citations counted in INSPIRE as of 19 Apr 2023

%\cite{Calibbi:2020jvd}
\bibitem{Calibbi:2020jvd}
L.~Calibbi, D.~Redigolo, R.~Ziegler and J.~Zupan,
``Looking forward to lepton-flavor-violating ALPs,''
JHEP \textbf{09}, 173 (2021)
doi:10.1007/JHEP09(2021)173
[arXiv:2006.04795 [hep-ph]].
%97 citations counted in INSPIRE as of 18 Apr 2023

%\cite{Belfatto:2019swo}
\bibitem{Belfatto:2019swo}
B.~Belfatto, R.~Beradze and Z.~Berezhiani,
``The CKM unitarity problem: A trace of new physics at the TeV scale?,''
Eur. Phys. J. C \textbf{80}, no.2, 149 (2020)
doi:10.1140/epjc/s10052-020-7691-6
[arXiv:1906.02714 [hep-ph]].
%99 citations counted in INSPIRE as of 19 Apr 2023

%\cite{Belfatto:2021jhf}
\bibitem{Belfatto:2021jhf}
B.~Belfatto and Z.~Berezhiani,
``Are the CKM anomalies induced by vector-like quarks? Limits from flavor changing and Standard Model precision tests,''
JHEP \textbf{10}, 079 (2021)
doi:10.1007/JHEP10(2021)079
[arXiv:2103.05549 [hep-ph]].
%35 citations counted in INSPIRE as of 15 Mar 2023

%\cite{Cheung:2020vqm}
\bibitem{Cheung:2020vqm}
K.~Cheung, W.~Y.~Keung, C.~T.~Lu and P.~Y.~Tseng,
``Vector-like Quark Interpretation for the CKM Unitarity Violation, Excess in Higgs Signal Strength, and Bottom Quark Forward-Backward Asymmetry,''
JHEP \textbf{05}, 117 (2020)
doi:10.1007/JHEP05(2020)117
[arXiv:2001.02853 [hep-ph]].
%29 citations counted in INSPIRE as of 15 Mar 2023

%\cite{Branco:2021vhs}
\bibitem{Branco:2021vhs}
G.~C.~Branco, J.~T.~Penedo, P.~M.~F.~Pereira, M.~N.~Rebelo and J.~I.~Silva-Marcos,
``Addressing the CKM unitarity problem with a vector-like up quark,''
JHEP \textbf{07}, 099 (2021)
doi:10.1007/JHEP07(2021)099
[arXiv:2103.13409 [hep-ph]].
%26 citations counted in INSPIRE as of 15 Mar 2023

%\cite{Botella:2021uxz}
\bibitem{Botella:2021uxz}
F.~J.~Botella, G.~C.~Branco, M.~N.~Rebelo, J.~I.~Silva-Marcos and J.~F.~Bastos,
``Decays of the heavy top and new insights on $\epsilon _K$ in a one-VLQ minimal solution to the CKM unitarity problem,''
Eur. Phys. J. C \textbf{82}, no.4, 360 (2022)
doi:10.1140/epjc/s10052-022-10299-9
[arXiv:2111.15401 [hep-ph]].
%5 citations counted in INSPIRE as of 19 Apr 2023

%\cite{Crivellin:2022rhw}
\bibitem{Crivellin:2022rhw}
A.~Crivellin, M.~Kirk, T.~Kitahara and F.~Mescia,
``Global fit of modified quark couplings to EW gauge bosons and vector-like quarks in light of the Cabibbo angle anomaly,''
JHEP \textbf{03}, 234 (2023)
doi:10.1007/JHEP03(2023)234
[arXiv:2212.06862 [hep-ph]].
%5 citations counted in INSPIRE as of 19 Apr 2023

%\cite{Belfatto:2023tbv}
\bibitem{Belfatto:2023tbv}
B.~Belfatto and S.~Trifinopoulos,
``The remarkable role of the vector-like quark doublet in the Cabibbo angle and $W$-mass anomalies,''
[arXiv:2302.14097 [hep-ph]].
%1 citations counted in INSPIRE as of 19 Apr 2023

%\cite{Fischer:2021sqw}
\bibitem{Fischer}
O.~Fischer, B.~Mellado, S.~Antusch, E.~Bagnaschi, S.~Banerjee, G.~Beck, B.~Belfatto, M.~Bellis, Z.~Berezhiani and M.~Blanke, \textit{et al.}
``Unveiling Hidden Physics at the LHC,''
[arXiv:2109.06065 [hep-ph]].
%6 citations counted in INSPIRE as of 30 Nov 2021 

%\cite{Anselm:1996jm}
\bibitem{MFVA}
A.~Anselm and Z.~Berezhiani,
``Weak mixing angles as dynamical degrees of freedom,''
Nucl. Phys. B \textbf{484}, 97-123 (1997)
doi:10.1016/S0550-3213(96)00597-4
[arXiv:hep-ph/9605400 [hep-ph]].
%33 citations counted in INSPIRE as of 31 Aug 2021

%\cite{Berezhiani:2001mh}
\bibitem{MFVR}
Z.~Berezhiani and A.~Rossi,
``Flavor structure, flavor symmetry and supersymmetry,''
Nucl. Phys. B Proc. Suppl. \textbf{101}, 410-420 (2001)
doi:10.1016/S0920-5632(01)01527-4
[arXiv:hep-ph/0107054 [hep-ph]].
%38 citations counted in INSPIRE as of 31 Aug 2021

%\cite{DAmbrosio:2002vsn}
\bibitem{MFVS}
G.~D'Ambrosio, G.~F.~Giudice, G.~Isidori and A.~Strumia,
``Minimal flavor violation: An Effective field theory approach,''
Nucl. Phys. B \textbf{645}, 155-187 (2002)
doi:10.1016/S0550-3213(02)00836-2
[arXiv:hep-ph/0207036 [hep-ph]].
%1627 citations counted in INSPIRE as of 31 Aug 2021

%%\cite{Zyla:2020zbs}
%\bibitem{PDG20}
%P.A.~Zyla \textit{et al.} [Particle Data Group], ``Review of Particle Physics,''
%PTEP \textbf{2020}, no.8, 083C01 (2020) %and 2021 update. 
%doi:10.1093/ptep/ptaa104

%\cite{Berezhiani:1998hg}
\bibitem{Berezhiani:1998hg}
Z.~Berezhiani, Z.~Tavartkiladze and M.~Vysotsky,
``d = 5 operators in SUSY GUT:  Fermion masses versus proton decay,''
[arXiv:hep-ph/9809301 [hep-ph]].

%%\cite{FlavourLatticeAveragingGroup:2019iem}
%\bibitem{FLAG19}
%S.~Aoki \textit{et al.} [Flavour Lattice Averaging Group],
%``FLAG Review 2019: Flavour Lattice Averaging Group (FLAG),''
%Eur. Phys. J. C \textbf{80}, no.2, 113 (2020)
%doi:10.1140/epjc/s10052-019-7354-7
%[arXiv:1902.08191 [hep-lat]].
%%517 citations counted in INSPIRE as of 02 Sep 2021

%\cite{FlavourLatticeAveragingGroupFLAG:2021npn}
\bibitem{FLAG21}
Y.~Aoki \textit{et al.} [Flavour Lattice Averaging Group (FLAG)],
``FLAG Review 2021,''
Eur. Phys. J. C \textbf{82}, no.10, 869 (2022)
doi:10.1140/epjc/s10052-022-10536-1
[arXiv:2111.09849 [hep-lat]].
%188 citations counted in INSPIRE as of 08 Nov 2022

%%\cite{Bazavov:2017lyh}
%\bibitem{MILC17}
%A.~Bazavov, C.~Bernard, N.~Brown, C.~Detar, A.~X.~El-Khadra, E.~G\'amiz, S.~Gottlieb, U.~M.~Heller, J.~Komijani and A.~S.~Kronfeld, \textit{et al.}
%``$B$- and $D$-meson leptonic decay constants from four-flavor lattice QCD,''
%Phys. Rev. D \textbf{98}, no.7, 074512 (2018)
%doi:10.1103/PhysRevD.98.074512
%[arXiv:1712.09262 [hep-lat]].
%%140 citations counted in INSPIRE as of 02 Sep 2021

%%\cite{MILC:2018ddw}
%\bibitem{MILC18}
%S.~Basak \textit{et al.} [MILC],
%``Lattice computation of the electromagnetic contributions to kaon and pion masses,''
%Phys. Rev. D \textbf{99}, no.3, 034503 (2019)
%%doi:10.1103/PhysRevD.99.034503
%[arXiv:1807.05556 [hep-lat]].
%%25 citations counted in INSPIRE as of 02 Sep 2021

%\cite{FermilabLattice:2018est}
\bibitem{Bazavov18}
A.~Bazavov \textit{et al.} [Fermilab Lattice, MILC and TUMQCD],
``Up-, down-, strange-, charm-, and bottom-quark masses from four-flavor lattice QCD,''
Phys. Rev. D \textbf{98}, no.5, 054517 (2018)
doi:10.1103/PhysRevD.98.054517
[arXiv:1802.04248 [hep-lat]].
%65 citations counted in INSPIRE as of 03 Sep 2021

%\cite{Colangelo:2018jxw}
\bibitem{Colangelo}
G.~Colangelo, S.~Lanz, H.~Leutwyler and E.~Passemar,
``Dispersive analysis of $\eta \rightarrow 3 \pi $,''
Eur. Phys. J. C \textbf{78}, no.11, 947 (2018)
doi:10.1140/epjc/s10052-018-6377-9
[arXiv:1807.11937 [hep-ph]].
%29 citations counted in INSPIRE as of 03 Sep 2021

%%\cite{Baikov:2016tgj}
%\bibitem{Chetyrkin5}
%P.~A.~Baikov, K.~G.~Chetyrkin and J.~H.~K\"uhn,
%``Five-Loop Running of the QCD coupling constant,''
%Phys. Rev. Lett. \textbf{118}, no.8, 082002 (2017)
%%doi:10.1103/PhysRevLett.118.082002
%[arXiv:1606.08659 [hep-ph]].
%%323 citations counted in INSPIRE as of 03 Sep 2021

%\cite{Chetyrkin:2000yt}
\bibitem{Chetyrkin}
K.~G.~Chetyrkin, J.~H.~Kuhn and M.~Steinhauser,
``RunDec: A Mathematica package for running and decoupling of the strong coupling and quark masses,''
Comput. Phys. Commun. \textbf{133}, 43-65 (2000)
%doi:10.1016/S0010-4655(00)00155-7
[arXiv:hep-ph/0004189 [hep-ph]].
%511 citations counted in INSPIRE as of 03 Sep 2021

%\cite{Machacek:1981ic}
\bibitem{Machacek0}
M.~E.~Machacek and M.~T.~Vaughn,
``Fermion and Higgs Masses as Probes of Unified Theories,''
Phys. Lett. B \textbf{103}, 427-432 (1981)
%doi:10.1016/0370-2693(81)90075-7
%77 citations counted in INSPIRE as of 10 Nov 2021

%\cite{Machacek:1983tz}
\bibitem{Machacek1}
M.~E.~Machacek and M.~T.~Vaughn,
``Two Loop Renormalization Group Equations in a General Quantum Field Theory. 1. Wave Function Renormalization,''
Nucl. Phys. B \textbf{222}, 83-103 (1983)
doi:10.1016/0550-3213(83)90610-7
%737 citations counted in INSPIRE as of 10 Nov 2021

%%\cite{Machacek:1983fi}
%\bibitem{Machacek2}
%M.~E.~Machacek and M.~T.~Vaughn,
%``Two Loop Renormalization Group Equations in a General Quantum Field Theory. 2. Yukawa Couplings,''
%Nucl. Phys. B \textbf{236}, 221-232 (1984)
%%doi:10.1016/0550-3213(84)90533-9
%%652 citations counted in INSPIRE as of 09 Sep 2021

%\cite{Luo:2002ey}
\bibitem{Luo}
M.~x.~Luo and Y.~Xiao,
``Two loop renormalization group equations in the standard model,''
Phys. Rev. Lett. \textbf{90}, 011601 (2003)
doi:10.1103/PhysRevLett.90.011601
[arXiv:hep-ph/0207271 [hep-ph]].
%156 citations counted in INSPIRE as of 09 Sep 2021

%\cite{Babu:1987im}
\bibitem{Babu}
K.~S.~Babu,
``Renormalization Group Analysis of the {Kobayashi-Maskawa} Matrix,''
Z. Phys. C \textbf{35}, 69 (1987)
doi:10.1007/BF01561056
%102 citations counted in INSPIRE as of 09 Sep 2021

%%\cite{Bednyakov:2012en}
%\bibitem{Bednyakov}
%A.~V.~Bednyakov, A.~F.~Pikelner and V.~N.~Velizhanin,
%``Yukawa coupling beta-functions in the Standard Model at three loops,''
%Phys. Lett. B \textbf{722}, 336-340 (2013)
%%doi:10.1016/j.physletb.2013.04.038
%[arXiv:1212.6829 [hep-ph]].
%%113 citations counted in INSPIRE as of 09 Sep 2021

%\cite{Barger:1992pk}
\bibitem{Barger}
V.~D.~Barger, M.~S.~Berger and P.~Ohmann,
``Universal evolution of CKM matrix elements,''
Phys. Rev. D \textbf{47}, 2038-2045 (1993)
doi:10.1103/PhysRevD.47.2038
[arXiv:hep-ph/9210260 [hep-ph]].
%78 citations counted in INSPIRE as of 09 Sep 2021

%%\cite{Ramond:1993kv}
%\bibitem{Ramond:1993kv}
%P.~Ramond, R.~G.~Roberts and G.~G.~Ross,
%``Stitching the Yukawa quilt,''
%Nucl. Phys. B \textbf{406}, 19-42 (1993)
%doi:10.1016/0550-3213(93)90159-M
%[arXiv:hep-ph/9303320 [hep-ph]].
%
%%\cite{Leurer:1993gy}
%\bibitem{Leurer:1993gy}
%M.~Leurer, Y.~Nir and N.~Seiberg,
%``Mass matrix models: The Sequel,''
%Nucl. Phys. B \textbf{420}, 468-504 (1994)
%doi:10.1016/0550-3213(94)90074-4
%[arXiv:hep-ph/9310320 [hep-ph]].
%
%%\cite{Berezhiani:1986xi}
%\bibitem{Berezhiani:1986xi}
%Z.~G.~Berezhiani and N.~I.~Shubitidze,
%``On the Structure of mass matrices of quarks,''
%Sov. J. Nucl. Phys. \textbf{44}, 357-358 (1986)

\end{thebibliography}
\end{document}